\documentclass[superscriptaddress,groupedaddress,nofootnoteinbib,11pt]{article}  

\usepackage{graphicx}  
\usepackage{dcolumn}   
\usepackage{bm}        
\usepackage{slashed}        
\usepackage{jcappub}
\usepackage{color}

\def\be{\begin{equation}}
\def\ee{\end{equation}}
\def\bea{\begin{eqnarray}}
\def\eea{\end{eqnarray}}
\def\beq{\begin{eqnarray}}
\def\eeq{\end{eqnarray}}

\def\p{{\cal P}}
\def\S{\Sigma}
\def\ep{\epsilon}

\def\L*{{\cal L}_*}
\def\L{\mathcal{L}}
\def\({\left(}
\def\){\right)}

\def\nn{\nonumber}
\def\p{\partial}

\def\p{\partial}

\def\<{\langle}
\def\>{\rangle}

 \def\neq {\not\equiv}

\def\VEV#1{\left\langle #1 \right\rangle}

\def\cs2{c_{s}^{2}}

 \def\ep{\epsilon}

 \def\p{\partial}

 \def\be   {\begin{equation}}   \def\ee   {\end{equation}}
 \def\bea  {\begin{eqnarray}}   \def\eea  {\end{eqnarray}}
 \def\bean {\begin{eqnarray*}}  \def\eean {\end{eqnarray*}}

\renewcommand{\thesubsection}{\arabic{section}.\arabic{subsection}}

\definecolor{RoyalBlue}{rgb}{0.25,.41,.88}
\definecolor{RedWine}{rgb}{0.743,0,0}

\begin{document}

\title{Inflationary tensor fossils in large-scale structure}

\author[a]{Emanuela Dimastrogiovanni,}
\author[b]{Matteo Fasiello,}
\author[c,d]{Donghui Jeong,}
\author[e]{Marc Kamionkowski}
\affiliation[a]{School of Physics and Astronomy, University of
     Minnesota, Minneapolis, MN 55455, USA}
\affiliation[b]{Department of Physics, Case Western Reserve University,
     Cleveland, OH 44106, USA}
\affiliation[c]{Department of Astronomy and Astrophysics, The
     Pennsylvania State University, University Park, PA 16802
     USA}
\affiliation[d]{Institute for Gravitation and the Cosmos, The
     Pennsylvania State University, University Park, PA 16802,
     USA}
\affiliation[e]{Department of Physics and Astronomy, 3400 N.\ Charles
     St., Johns Hopkins University, Baltimore, MD 21218 USA}
\emailAdd{ema@physics.umn.edu}
\emailAdd{mrf65@case.edu}
\emailAdd{kamion@jhu.edu}
\emailAdd{duj13@psu.edu}

\abstract{Inflation models make specific predictions for a
tensor-scalar-scalar three-point correlation, or bispectrum,
between one gravitational-wave (tensor) mode and two
density-perturbation (scalar) modes.  This tensor-scalar-scalar
correlation leads to a local power quadrupole, an apparent
departure from statistical isotropy in our Universe, as well as
characteristic four-point correlations in the current mass
distribution in the Universe.  So far, the predictions for these
observables have been worked out only for single-clock models in
which certain consistency conditions between the
tensor-scalar-scalar correlation and tensor and scalar power
spectra are satisfied. Here we review
the requirements on inflation models for these
consistency conditions to be satisfied.  We then consider
several examples of inflation models, such as non-attractor
and solid-inflation models, in which these conditions are put to the test.
In solid inflation the simplest consistency conditions are already violated whilst in the non-attractor model we find that, contrary to the standard scenario, the tensor-scalar-scalar correlator probes directly relevant model-dependent information.
We work out the predictions for observables in these
models.  For non-attractor inflation we find an apparent
local quadrupolar departure from statistical isotropy in
large-scale structure but that this power quadrupole decreases
very rapidly at smaller scales.  The consistency of the CMB
quadrupole with statistical isotropy then constrains the
distance scale that corresponds to the transition from the
non-attractor to attractor phase of inflation to be larger than
the currently observable horizon.  Solid inflation predicts
clustering fossils signatures in the current galaxy distribution
that may be large enough to be detectable with forthcoming, and
possibly even current, galaxy surveys.}
\leftline{FTPI-MINN-14/21}\leftline{UMN-TH-3346/14}
\maketitle

\section{Introduction}

The notion that the Universe began with a period of inflationary
expansion
\cite{Guth:1982ec,Bardeen:1983qw,Hawking:1982cz,Linde:1981mu,Mukhanov:1981xt}
has been supported with a series of increasingly precise
observational tests.  The tests have verified that a number of
characteristics of primordial density (scalar metric)
perturbations---including adiabaticity, Gaussianity, and near
scale-independence---agree with those in these simplest
single-field slow-roll (SFSR) models.  Still, inflation has even
more consequences, including the prediction of a nearly
scale-invariant spectrum of primordial gravitational waves
\cite{Abbott:1984fp,Rubakov:1982df,Fabbri:1983us,Starobinsky:1979ty}.
Even though it is too early to attribute conclusively the CMB
B modes \cite{Kamionkowski:1996zd,Seljak:1996gy} detected by
BICEP2 \cite{Ade:2014xna} to inflationary gravitational waves,
evidence for a scalar spectral index $n_s\neq1$
\cite{Hinshaw:2012aka,Ade:2013zuv} provides some reason to believe
that the gravitational-wave amplitude might be large.
It is thus warranted to think about other obervational
consequences of primordial gravitational waves and what we can
learn about inflation from such observations.

In particular, there is a growing body of work on the imprints
of gravitational waves on large-scale structure.  Lensing by
tensor modes of the galaxy distribution
\cite{Dodelson:2003bv,Schmidt:2012nw,Dai:2012bc,Chisari:2014xia}, CMB
\cite{Cooray:2005hm,Li:2006si,Dodelson:2010qu,Book:2011na}, and
21-cm fluctuations \cite{Pen:2003yv,Masui:2010cz,Book:2011dz}
have been studied.  However, there may also be imprints on the
unlensed mass distribution.  One possibility is that
long-wavelength gravitational waves may give rise to an apparent
local departure from statistical isotropy in the form of a power
quadrupole
\cite{Giddings:2011zd,Dai:2013ikl,Dai:2013kra,Brahma:2013rua}
that can be observed in the CMB
\cite{Pullen:2007tu,Hanson:2009gu} and large-scale structure
\cite{Ando:2008zza}; some (null) CMB
\cite{Groeneboom:2008fz,Bennett:2010jb,Ade:2013nlj} and
large-scale-structure \cite{Pullen:2010zy} searches 
have already been carried out.  There are then higher-order
correlations in the density distribution, induced by coupling to
gravitational waves, that can be sought
\cite{Seery:2008ax,Masui:2010cz,Jeong:2012df,Jeong:2012nu,Schmidt:2012nw,Schmidt:2013gwa}
and also possible signatures
\cite{Schmidt:2012nw,Schmidt:2013gwa} in the tidally-induced
intrinsic alignments of galaxies \cite{Catelan:2000vm}.

The purpose of this paper is to study the prospects to learn
about inflation through the tensor-scalar-scalar (TSS) correlation.
Such a correlation arises even in the simplest SFSR models
\cite{Maldacena:2002vr}.  This SFSR tensor-scalar-scalar bispectrum
satisfies a particular consistency condition (cc) that relates
the functional dependence of the TSS bispectrum on the tensor
wavenumber $K$ and scalar wavenumbers $k_1$ and $k_2$  to the
tensor and scalar power spectra in the squeezed limit ($K \ll
k_1,k_2$)
\cite{Maldacena:2002vr,Sreenath:2013xra,Sreenath:2014nka}.
This TSS consistency condition parallels an
analogous consistency condition for the scalar three-point
function (SSS) that is known to hold not just for SFSR
inflation, but for single-clock models more generally
\cite{Creminelli:2004yq,Kehagias:2012pd,Creminelli:2012qr,Goldberger:2013rsa,Hinterbichler:2013dpa,Kehagias:2013xga,Berezhiani:2013ewa,Berezhiani:2014tda}. As
we show below, arguments like those for the generality of the
SSS consistency condition apply also to the TSS consistency
condition, and so the TSS consistency condition is generic
to a fairly broad class of inflationary models.

The primordial TSS bispectrum that arises if
the consistency condition is satisfied naively implies a power
quadrupole that suffers a logarithmic (for a scale-independent
tensor power spectrum) infrared divergence
\cite{Jeong:2012df,Dai:2013ikl}.  However, that divergence is
cancelled precisely by an analogous divergence at late times in 
the projection effect including lensing deflection \cite{Senatore:2012nq,Pajer:2013ana,Dai:2013kra} leaving
a finite observable power quadrupole
\cite{Dai:2013kra,Schmidt:2013gwa}, which although nonzero turns
out to be quite small.  The cancellation arises, though, only if
the consistency condition is satisfied.  Thus, if a power
quadrupole in excess of 
that expected from this consistency condition were to be
discovered, it would rule
out the vast majority of single-field inflationary models.
It is therefore crucial to understand the origin, the scope, and
the limits of the TSS consistency condition in inflationary setups.

Below, in Section~\ref{sec:origin}, we first review the origin
of the scalar-scalar-scalar cc and then how, on general grounds, the TSS consistency condition arises analogously.    In Section
\ref{sec:nonattractor} we then consider non-attractor
inflation \cite{Namjoo:2012aa,Chen:2013eea,Kinney:2005vj}.
These models consist nominally of only a single scalar
field.  However, the slow-roll phase that is
assumed to have been reached in SFSR models is not
satisfied.  The violation in these models of the slow-roll
conditions implies that the scalar-field equation of motion is a 
second-order differential equation rather than a first-order
differential equation.  The physical conditions in the Universe
at some particular time are therefore not determined exclusively
by the inflaton field value, and so the single-clock conditions
may be violated.  We calculate the TSS
bispectrum and show that, intriguingly, it directly probes model-dependent information.  In Section
\ref{sec:solid} we consider solid inflation \cite{Endlich:2012pz}
as another example of a model in which the consistency condition
may be violated.  In Section \ref{sec:observations} we calculate
the observable power quadrupole in these models and also study the
prospects to discern from higher-order galaxy clustering the
differences between the TSS in these models.  We then conclude
in Section \ref{sec:conclusion}.

\section{Scalar-scalar-scalar and tensor-scalar-scalar
consistency conditions}
\label{sec:origin}

A great deal of recent attention
\cite{Creminelli:2004yq,Kehagias:2012pd,Creminelli:2012qr,Goldberger:2013rsa,Hinterbichler:2013dpa,Berezhiani:2013ewa,Berezhiani:2014tda}
has been  directed towards clarifying the origin of the
consistency conditions for cosmological correlators \cite{Maldacena:2002vr}. 
The scope of the consistency conditions extends well beyond
single-field slow-roll inflation \cite{Creminelli:2004yq}.
However, as exemplified in
Refs.~\cite{Namjoo:2012aa,Chen:2013eea} (see also
Ref.~\cite{Kinney:2005vj}), violations can occur already in
single-field inflation.  As we will see, these models share a
specific property that causes them to evade the consistency
conditions.\footnote{This is true at least at lowest order in
the soft momentum $q$, in their simplest and model-independent
formulation.}

The key realization is that the consistency conditions are a
direct consequence of
the invariance under space
diffeomorphisms of the classical and quantum theory. There is
more to it: crucially one need not necessarily specify the form
of the action as long as the symmetry is in place and a
\textit{locality} requirement\footnote{The generic locality
requirement is equivalent to \textit{adiabaticity} in
cosmological parlance \cite{Hinterbichler:2013dpa}.} is
satisfied. This latter realization is at the heart of the
generalization of the results of Ref.~\cite{Maldacena:2002vr}
found in Ref.~\cite{Creminelli:2004yq} and put on firmer ground
by Ref.~\cite{Hinterbichler:2013dpa}. 

Whenever the locality requirement is met, the consistency
conditions for a scalar (tensor) with two hard scalars at zeroth
order in the soft momentum $q$ take the familiar form,
\bea
     \frac{\<\zeta_{\vec q}\zeta_{\vec p}\zeta_{-\vec q-\vec
     p}\>}{P_{\zeta}(q)}= -\left(3+p_i \frac{\p}{\p p_i} \right)
     P_{\zeta}(p);\qquad \frac{\<\gamma^{ij}_{\vec q}
     \zeta_{\vec p} \zeta_{-\vec q-\vec p}\>}{P_{\gamma}(q)} =
     -\frac{1}{2}\hat{P}^{ijkl}(\hat{q}) p_k\frac{\p}{\p p_l}
     P_{\zeta}(p), \nonumber\\ \label{simple}
\eea
where $q \ll p$ is the wavenumber of the long-wavelength mode,
$p$ that of the short-wavelength modes, and $P_\zeta(k)$ and
$P_\gamma(k)$, respectively, the scalar and tensor power
spectra defined as 
\bea
\left<
\zeta({\vec k})
\zeta({\vec k}')
\right>
=
(2\pi)^3 P_{\zeta}(k)\delta^D({\vec k} + {\vec k}')
\\
\left<
\gamma_{ij}^\lambda({\vec k})
\gamma_{ij}^\lambda({\vec k}')
\right>
=
(2\pi)^3 4 P_{\gamma}(k)\delta^D({\vec k} + {\vec k}').
\eea

In words, this
specific limit of a three-point function depends entirely on the
behavior of the two-point correlator, regardless of the
information content (i.e. interactions) one can access only at
the level of the cubic action.

We give below a brief account (based mainly on \cite{Berezhiani:2013ewa}) of how cosmological consistency
conditions such as the ones above are derived to all orders in
the soft momentum $q$ as a consequence of the Slavnov-Taylor
identity for spatial diffeomorphisms. For a thorough treatment,
we refer the reader to
Refs.~\cite{Berezhiani:2013ewa,Goldberger:2013rsa,Berezhiani:2014tda}.

\subsection{Origin of the ccs}

We use the fixed-time path integral of
Ref.~\cite{Goldberger:2013rsa}, a 3-D Euclidean path integral
over configurations at the final time, with the wavefunction
storing the ``history'' information. The fluctuations around an
FRW background are,
\bea
     g_{\mu\nu}=\bar{g}_{\mu\nu}+a^2(t) h_{\mu\nu}={\rm
     diag}(-1, a^2(t))+a^2(t) h_{\mu\nu}\,;\quad
     \phi(x,t)=\bar{\phi}(t)+ \varphi(t,x).
\eea
Fixed-time correlators, to be turned into cosmological
observables, can be obtained from the generating functional,
\bea
     Z[T,J]=\int \mathcal{D}h_{ij}\mathcal{D} \varphi
     |\Psi[h,\varphi,t]|^2 {\rm Exp}\left(\int d^3 x (h_{ij}
     T^{ij}+\varphi J) \right),
     \label{ZZ}
\eea
where $J$ and $T$ are the currents and $\Psi$ the wave function.

Since the theory involves general relativity
plus a scalar field, both the action and the functional measure
will be invariant under space diffeomorphisms (we have
surrendered time-reparametrization invariance
by choosing the fixed-time formalism).  Requiring that the
same is true for the generating functional ($Z$ ought to be
invariant under a field-redefinition) amounts to the condition
\cite{Berezhiani:2013ewa},
\bea
\nonumber
     0 & = & \int \mathcal{D}h_{ij} \mathcal{D}\varphi
     \left\vert \Psi[h,\varphi, t]\right\vert^2 e^{S_{\rm
     ource}} \int {\rm d}^3x~ \xi^k\bigg\{\text{(G.F.)}_k -2
     \p_j T^{j}_{k}+\p_k h_{ij} T^{ij}-2\p_j \left( h_{ik}
     T^{ij} \right)+\p_k \varphi J \bigg\} \\ \nonumber
     &=&  \int {\rm d}^3x~ \xi^k \left\{ \text{(G.F.)}_k-2 \p_j
     T^{j}_k+\p_k \left(\frac{\delta}{\delta T^{ij}}\right)
     T^{ij}-2\p_j \left(\frac{\delta}{\delta T^{ik}} T^{ij}
     \right)+\p_k \left(\frac{\delta}{\delta J}\right) J
     \right\}Z[T,J],
\label{zetaVar}
\eea
which is obtained by varying the gauge-fixing term and the
source term within $Z$ under space diffeomorphisms,
\bea
     \varphi\rightarrow \varphi+\xi^k \p_{k}\varphi ; \quad
     h_{ij}\rightarrow h_{ij}+\p_i \xi_j+\p_j \xi_i+\xi^k \p_{k}
     h_{ij} + h_{jk}\p_i \xi^k+h_{ik}\p_j \xi^k,
\eea
and where $S_{ource}$ stands for the argument of the exponential in Eq.~(\ref{ZZ}).
The term $\text{(G.F.)}_k$ denotes the contribution
originating from the variation of the gauge-fixing
term.  Although essential for the general formula, this
contribution plays no role in the derivation of tree-level
consistency conditions for a soft tensor with two hard scalar
modes that we are after.

Our starting point here has been a functional integral over the
metric entries $h_{ij}$ only, while $h_{00}$ and $h_{0i}$ have
been already integrated out by solving the equation of
motion. This  will be essential in a few steps.  Indeed, despite
being handed a theory which is clearly local, integrating out
$h_{00(i)}$ (alternatively known as $N_1,N_i$ in the ADM
formalism and in Ref.~\cite{Maldacena:2002vr}) results in
spatially non-local terms (recall that, e.g., $N_i\sim \p_i
\p^{-2}(\ldots)$).

From the second line in Eq.~(\ref{zetaVar}) one can see that, $\xi$
being arbitrary, the entire integrand must vanish as a result of
the invariance of $Z$ under field redefinitions. It is
thus convenient to introduce the effective action,
\bea
     \Gamma[h,\varphi]=\ln Z -\int d^3 x (h_{ij}T^{ij}+\varphi J) .
\eea
Upon choosing to work in comoving gauge,\footnote{The comoving
gauge is defined by $\quad \varphi=0\,; \quad
\delta_{ij}+h_{ij}=e^{2\zeta}\hat{h}_{ij}\,;\qquad  {\rm with}
\qquad {\rm det}\,\hat{h}=1\,; \quad \gamma_{ij}\equiv {\rm
ln}\,\hat{h}_{ij}\,; \quad \gamma^{i}_i=0 $}\ requiring a
vanishing integral amounts to an identity of the form,
\bea
     \frac{1}{\alpha}\left( \vec{\nabla}^2 \p_j
     \gamma_{ij}+\p_i\p_j\p_k \gamma_{jk}
     \right)+2\p_j\left(\frac{1}{6}\,\delta_{ij}\,\frac{\delta\Gamma}{\delta
     \zeta}+\frac{\delta\Gamma}{\delta \gamma_{ij}} \right)=\p_i
     \zeta \frac{\delta\Gamma}{\delta \zeta}+\cdots,
\eea  
where the dots denote contributions that are higher order in
$\gamma$ and thus irrelevant for the
tree-level identity. The term proportional to $\alpha^{-1}$,
originating from the gauge-fixing part, will not play a role in
what follows. 

To obtain the consistency conditions in
their more familiar form, we note that the effective action
$\Gamma$ is the generating function of the 1PI
correlation functions.  Since we are after the squeezed limit of
a three-point function with a soft tensor (scalar) mode and two
hard scalars, it suffices to differentiate the expression above
twice with respect to $\zeta$.  In Fourier space,
\bea
     \frac{1}{3} q_i \Gamma^{\zeta\zeta \zeta}
     (\vec q, \vec p, -\vec q -\vec p)+2q^j
     \Gamma_{ij}^{\zeta\zeta \gamma} (\vec q, \vec p, -\vec q
     -\vec p)=q_i \Gamma_{\zeta}(p)-p_i\left(
     \Gamma_{\zeta}(|\vec q +\vec p|)-\Gamma_{\zeta}(p)
     \right). \nonumber \\
\label{main}
\eea
where $\Gamma^{\zeta\zeta\zeta}$, $\Gamma^{\zeta\zeta\gamma}$, 
$\Gamma_\zeta$ are functional derivatives of the effective
action $\Gamma$ with respect to their indices 
(e.g., $\Gamma_\zeta\equiv\delta^2\Gamma/\delta\zeta^2$), 
which are proportional to, respectively, $\left<\zeta\zeta\zeta\right>$,
$\left<\zeta\zeta\gamma\right>$, and $\left<\zeta\zeta\right>$.
The solution to Eq.~(\ref{main}) can be obtained
\cite{Berezhiani:2013ewa} order by order in $q$.  Once a
suitable projection operator $\hat P$ has been found
\cite{Berezhiani:2013ewa}, the
solutions for the individual vertices are,
\begin{equation}
     \frac{\< \zeta_{\vec{q}} \zeta_{\vec{p}}
     \zeta_{-\vec{q}-\vec p} \>}{P_{\zeta}(q)}=K(\vec p, \vec
     q)+A(\vec p, \vec q), \quad
     \frac{\< \gamma^{ij}_{\vec q} \zeta_{\vec{p}}
     \zeta_{-\vec{q}-\vec p}
     \>}{P_{\gamma}(q)}=\frac{1}{2}\hat{P}^{ijkl}(\hat q)
     (K_{kl}+A_{kl}),
\label{main2}
\end{equation}
where $K_{ij}$ is an array completely determined out of $P_{\zeta}$
while, most importantly for us, $A_{ij}$ is an
\textit{arbitrary} symmetric and transverse matrix ($K,A$ being the respective traces). The latter,
in full generality, will be of the form,
\bea
     A_{ij}= \epsilon_{ikm} \epsilon_{jlm} q^k q^l \left( a(\vec
     q,\vec p)\delta^{mn}+b(\vec q,\vec p)p^m p^n \right)
     . \label{adef}
\eea 
Thus, exploiting one's knowledge on the symmetry of the action
can prove extremely useful and deliver a relation between an
$N+1$ and an $N$-point correlator.  But, before a crucial
hypothesis in made on $A_{ij}$, it can only take us so
far. Indeed the physics that is specific to the dynamics of a
given inflationary phase (let alone an inflationary model) is
encoded into the functions $a,b$.

It seems reasonable to assume that $A$ starts out as a quadratic
function in $q$, which would be tantamount to saying that $a,b$ are local
functions. It would imply that there is an order $q^0$ and an
order $q$ consistency condition to be derived without the need
of the explicit information stored in $a,b$. If this were indeed
the case, the $q^0$-order ccs for a soft scalar and a soft
tensor would read as in Eq.~(\ref{simple}), a result one can
arrive at from Eq.~(\ref{main2}) using the explicit solution for
$K_{ij}$.\footnote{We refer the interested reader to
Ref.~\cite{Berezhiani:2013ewa} for an explicit expression. The crux
of the matter here is that $K_{ij}$ is entirely 
determined from one's knowledge of the 2-point function only.}

Crucially, as we stressed above, the theory one deals with at
this point follows from integrating out the lapse and the shift
function using the constraint equations. In particular, the
shift is,
\bea
     N_i\sim \epsilon\, \frac{q_i}{q^2}\,\dot\zeta+{\rm local}.
\eea
Now, since the action (and therefore the three-point function)
is also composed of terms like the non-local contribution
above, the only way to be sure that this will not result in a
non-analytic $a,b$ is to check a non-trivial property on the $\dot
\zeta$ part, namely that its time derivative is such that this
contribution becomes local; e.g., $\dot \zeta \sim q^2$.

This is precisely what happens whenever $\zeta$ is an adiabatic
mode \cite{Weinberg:2003sw}, and is therefore conserved outside the
horizon.  Conversely, as soon as $\zeta$ is not conserved, $A$
might well play a role in the consistency conditions already at order one or
zero. But those orders in $q$ are exactly the ones that gave
us\footnote{The $q^1$ part of Eq.~(\ref{simple}) has been omitted
in the Introduction but can be found in \textit{Appendix
A}.}\ Eq.~(\ref{simple}) which we now recognize as valid only
for a class, albeit a very populated one, of possible cases.

\subsection{Intuitive understanding of the CCS}

We now provide a more intuitive understanding of how and why the
ccs do not take the usual form, Eq.~(\ref{simple}), for cases
such as the one of a non-attractor solution. The familiar
intuitive picture is due to Maldacena \cite{Maldacena:2002vr}:
the effect of e.g. a soft scalar mode on two hard scalars is
that of anticipating their horizon exit by an amount $\delta
t^{*}$ which is easily calculated as follows,
\bea
     e^{2\zeta_S}e^{2\zeta_L}a^2(t)\simeq
     (1+2\zeta_L)e^{2\zeta_S} a^2(t)\sim e^{2\zeta_S}
     a^2(t+\delta t^{*})\,\,\,\rightarrow\,\,\, \delta t^{*}\sim
     -\zeta_L/H,
\label{intuitivesss}
\eea
so that the overall effect on the hard modes is well
approximated by $-\zeta_L/H\times\, d/dt^{*}\left(
\<\zeta_S\zeta_S \>\right)$, and the overall squeezed limit of
the scalar three-point function is proportional to the
long-wavelength spectrum times the tilt of the short-wavelength
spectrum. An analogous relation is found along similar lines
whenever the soft mode is a tensor one. 

Eq.~(\ref{intuitivesss}) is a good approximation because one is
expanding the exponential at a time around $t^*$ when the mode
$\zeta_L$ has long left the horizon and does
not depend on time anymore.  If this latter condition were not
satisfied, one would have to keep track of the history of $\zeta_L$
since it left the horizon. 
 
Ref.~\cite{Creminelli:2004yq} present a case for consistency
conditions in single-clock solutions whose background is a dynamical
attractor.  In non-attractor inflation, the evolution of the
inflaton has not yet reached the attractor phase---the decaying
mode has not yet decayed away.  The physical conditions at some
point in the Universe therefore cannot, as in single-clock
inflation, be specified entirely by a single parameter.

The perspective of Ref.~\cite{Creminelli:2004yq} is to reabsorb
$\zeta_L$ into the comoving \textit{space} variable---$i.e.$, in
the metric $e^{2\zeta_L} d{\bf x}^2=d{\bf \tilde{x}}^2$---so that
one can then effectively trade $d/d\zeta_L$ with $x\cdot d/dx$
in calculating the effect of a soft modes on two hard
ones.\footnote{The difference, as well as the overall analogy
with Ref.~\cite{Maldacena:2002vr}, where $\zeta_L$ is embedded into
the scale factor, is clear.}\ This trading evidently becomes
off limits as soon as $\zeta_L$ is time dependent thus taking us
back to the locality (or adiabaticity) argument discussed
above.

\subsection{Preview of non-attractor inflation}
\label{preview}
Before detailing its dynamics,  we give a qualitative picture of
the non-attractor inflationary phase from the perspective of ccs
violation. As shown in Ref.~\cite{Chen:2013eea}, the observable
quantity $\< \zeta \zeta \zeta\> $ violates the ccs already at
order $q^0$. As is clear from Eq.~(\ref{adef}) then, this can
only \footnote{$K_{ij}$ is always a local function.}\ result
from non-locality either in the function $a(q,p)$ or in the
function $b(q,p)$.

What we are after here is instead the consistency relation (or
violation thereof) of the observable $\<\gamma \zeta
\zeta\>$. As a result, the role of the projection operator
$\hat{P}^{ijkl}$ in Eq.~(\ref{main2}) is of the utmost
importance. In might indeed happen that $\hat{P}$ projects out
the non-analyticity of $a$ or $b$ from contributing to the
tensor-scalar-scalar three-point function at some (or all)
orders in $q$.  As it turns out, this is indeed what happens for
the non-attractor model under scrutiny in this
manuscript and therefore, in this strict sense, one should say that the TSS three-point function does not violate the consistency conditions. Nevertheless, as we shall see, one is able to probe the model-dependent part of the inflationary mechanism encoded in $A_{ij}$ by accounting for the contribution of
interaction terms which are simply slow-roll suppressed in the typical
attractor models.

The key point is that, in the non-attractor phase, the small
parameter $\epsilon$ is strongly time dependent, and it is
precisely this time dependence that can compensate for the
slow-roll suppression. The three-point-function contributions of
these interactions scale like $q^2$  and are therefore ideal
candidates to probe the content of $A_{ij}$, be it the non-local
or the analytic/adiabatic one. We will see below how there are
clear-cut ways to discern whether these $q^2$ terms contribute
to $A_{ij}$, or instead to the model-independent piece
$K_{ij}$. 

In all that follows, the characterization of the TSS signal of the non-attractor model as \textit{violating} the ccs is to be understood only in the sense above, that is: the squeezed limit is probing model-dependent information of the non-attractor model through contributions to the three-point function which are not naively slow-roll suppressed. The TSS carries an imprint of the non-attractor phase that cannot be extracted from the power spectrum nor from its derivatives. This is in contradistinction to the vast majority of inflationary models, including SFSR.

\begin{figure}
\begin{center}
\includegraphics[scale=1.0]{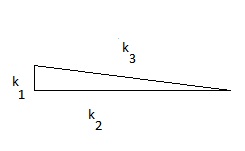}
\caption{Squeezed configuration for the momenta: $k_{1}\ll k_{2}\simeq k_{3}$. }
\label{fig1}
\end{center}
\end{figure}

\section{Inflation with a Non-Attractor Phase}
\label{sec:nonattractor}

As discussed above, non-attractor models are those
where the decaying solution for the inflaton equation of motion
has not yet fully decayed.  Following previous work on
non-attractor inflation, we consider \textsl{k-inflation}
\cite{ArmendarizPicon:1999rj} models described by a Lagrangian
density $P(X,\phi)$ that is a function of the inflaton $\phi$
and $X\equiv -1/2(\partial_{\mu}\phi)^2$.  The model is
parametrized by the quantities,
\begin{equation}
     c_{s}^{2} \equiv \frac{P_{,X}}{P_{,X}+2XP_{,XX}}, \quad
     \epsilon  \equiv
     -\frac{\dot{H}}{H^{2}}=\frac{XP_{,X}}{M_{P}^{2}H^{2}}  ,\quad
     \eta \equiv
     \frac{\dot{\epsilon}}{H\epsilon} =
     \frac{\ddot{\phi}}{H\dot{\phi}}
     \left(1+\frac{1}{c_{s}^{2}}\right) +
     \frac{\dot{\phi}P_{,X\phi}}{HP_{,X}}+2\epsilon.
\end{equation}
The sound speed $c_{s}$ varies between $0$ and $1$.
The slow-roll parameters $\epsilon$ and
$\eta$ are generally small and vary slowly with time, although
these assumptions are modified during the non-attractor
phase. 

\begin{figure}
\begin{center}
\includegraphics[scale=0.57]{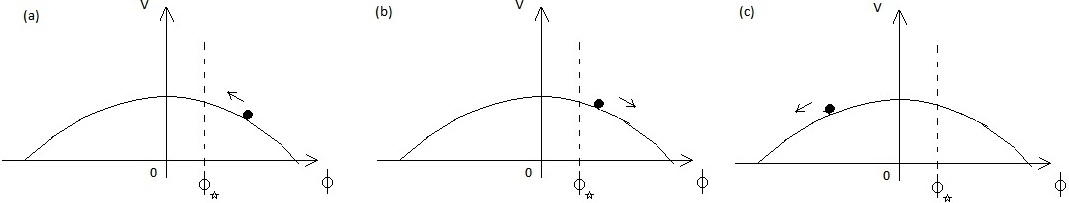}
\caption{A qualitative representation in the potential-scalar
     field plane of the non-attractor phase (a), and of
     the attractor phase, in the form of the (b) undershoot and
     overshoot (c) cases. The scalar field begins with a positive
     value. In the undershoot case, the field reaches a value
     $\phi_{*}>0$ at the point where its velocity becomes null,
     so it rolls down on the same side of the potential (b). In
     the overshoot case, the system has enough kinetic energy to
     go over the top of the potential and roll
     down the other side (c).}
\label{zero}
\end{center}
\end{figure}

Ref.~\cite{Chen:2013eea} presents a model of non-attractor
inflation that involves a Lagrangian density,
\begin{equation}\label{true}
     P(X,\phi)=X+\frac{X^{\alpha}}{M^{4(\alpha -1)}} - V(\phi),
     \quad\quad\quad\quad
     V(\phi)=V_{0}+v\left(\frac{\phi}{M_{P}}\right)^{\beta},
\end{equation}
where $\alpha$, $V_{0}$, $v$, and $\beta$ are initially free
parameters and $M_{P}\equiv 1/\sqrt{8\pi G_{N}}$ is the reduced Planck mass. 
During the first phase of inflation, the system evolves around a
non-attractor background: the inflaton rolls up its
potential, progressively slowing down (part~(a) of
Fig(\ref{zero})). This first phase can be followed by an
attractor inflationary phase with the inflaton field rolling
back down its potential on the same $\phi>0$ side (``undershoot
case'', see part~(b) of Fig.~\ref{zero}) or alternatively by it
going over the top of the potential and rolling down on the
opposite side (``overshoot case'', part~(c) of
Fig.~\ref{zero}).

For inflation to occur, the constant contribution $V_{0}$ to the
potential must dominate the total energy density during
inflation; i.e., $3H^{2}M_{P}^{2}\simeq V_{0}$.  The
kinetic energy of the inflaton must initially be large enough to
induce the field to roll up; if one considers values for
the exponent $\alpha\gg 1$ in Eq.~(\ref{true}), then large
values of $|\dot{\phi}|$ imply that the kinetic Lagrangian is
initially dominated by the non-canonical term,
$X^{\alpha}/M^{4(\alpha-1)}$.  Having a  large value for
$\alpha$, together with  $X^{\alpha}/M^{4(\alpha-1)}\gg X$,
results in a sound speed, $c_{s}^{2}\simeq
(2\alpha-1)^{-1}\lesssim 1$.  Following
Ref.~\cite{Chen:2013eea}, we choose $\alpha>1$.  To simplify, we
assume that both $\eta$ and $c_{s}$ are approximately 
constant during inflation. If $\eta$ is indeed a constant, the
other slow-roll parameter $\epsilon$ would go like $\epsilon\sim
a^{\eta}$ so that an $\eta\ll 1$ would result in an
approximately time-independent $\epsilon$ during
inflation. However, as we will see from the analysis of linear
perturbations, near scale invariance of the power spectrum in
this model requires a large $\eta\simeq -6$, and hence a
non-negligible time dependence for $\epsilon$. 
Ref.~\cite{Chen:2013eea} considered the ansatz $\phi\sim
a^{\kappa}$ for the homogeneous evolution of the inflaton, where
$\kappa$ is a constant. One then finds,
\begin{equation}\label{parone}
     \beta =  2\alpha=1+\frac{1}{c_{s}^{2}},\quad
     v =
     -\frac{M^{4}}{c_{s}^{2}} \left(\frac{V_{0}\kappa^{2}}{
     6M^{4}}
     \right)^{\alpha}\left(1+\frac{3c_{s}^{2}}{\kappa}\right), \quad
     \epsilon \sim a^{2\alpha\kappa} \quad\rightarrow\quad
     \kappa\simeq\frac{\eta}{2\alpha}.
\end{equation}
For given values of $\eta$ and $c_{s}$, we are
left with two independent parameters; e.g., $V_0$ and $v$. As
anticipated, $\eta$ is fixed by scale invariance while the
value of $c_{s}$ is essential to determine the amplitude of
non-Gaussianity. Notice that $v<0$, so the potential $V(\phi)$
has a concave shape as in Fig.~(\ref{zero}). Also $\kappa<0$
and, as a consequence, both $|\phi|$ and $|\dot{\phi}|$ are
decreasing functions of time during this initial phase. 
The {\it ansatz} $\phi\sim a^\kappa$ and the constraints on the
parameters derived from it represent an analytic solution that
complies with the dynamics of a non-attractor phase, 
but they are not a necessary condition for a non-attractor phase.

In the undershoot situation, the field stops before reaching
zero value.  Therefore, as we approach the turnaround point
($\phi_{*}$), the $X^{\alpha}$ contribution to the kinetic
energy becomes progressively negligible compared to $X$. The
condition that defines the end of the non-attractor phase is
$(\phi_{*}/M_{P}) \simeq (\sqrt{6}/|\kappa|)( M^{2}/\sqrt{V_{0}}).$
For $t>t_{*}$ the system transitions to
slow-roll inflation during which the curvature perturbation is
conserved on superhorizon scales. 

\subsection{Review of linear perturbations}

We decompose the metric fluctuations using the ADM
formalism \cite{Arnowitt:1962hi},
\begin{equation}\label{admf}
     ds^{2}=-N^{2}dt^{2}+h_{ij} \left(dx^{i}+N^{i}dt\right)
     \left(dx^{j}+N^{j}dt\right).
\end{equation}
We work in comoving gauge, setting to zero the fluctuations of
the inflation field ($\delta\phi=0$). In this gauge, the spatial
part of the metric tensor has the form, $h_{ij} = a^{2}
e^{2\zeta} \left(e^{2\gamma}\right)_{ij}$,
where $a(t)$ is the scale factor, $\zeta(t,\vec{x})$ is the
scalar fluctuation, and $\gamma_{ij}(\vec{x},t)$ is the
transverse traceless tensor perturbation (with
$\partial_{i}\gamma_{ij}=\gamma_{ii}=0$).  

Following the usual treatment, the mode function for the Fourier
component $\zeta_k(\tau)$ during the attractor phase is
\begin{equation}\label{Att}
     \zeta_{k} = \frac{v_{k}}{z}=\frac{H_{*}}{M_{P}}
     \frac{e^{-ikc_{s}\tau}}{\sqrt{4\epsilon_{*}c_{s}k^{3}}}
     \left(1+ikc_{s}\tau\right).
\end{equation}
In the non-attractor case, the time variation of $\epsilon$
(recall $\epsilon\sim a^{-6}$) leads to a different mode
function \cite{Chen:2013eea},
\begin{equation}\label{ratt}
     \zeta_{k}(\tau) = C_{k}\sqrt{\frac{2}{\pi}}
     \frac{e^{-ikc_{s}\tau}}{(-kc_{s}\tau)^{3}}
     \left(-1-ikc_{s}\tau\right),  \quad\quad\quad\quad
     |C_{k}|^{2}\equiv\frac{H_{*}^{2}}{M_{P}^{2}}
     \left(\frac{\pi}{8}\right) \frac{k^{3}c_{s}^{
     5}\tau_{*}^{6}}{\epsilon_{*}},
\end{equation}
where a $^{``\,*\,"}$ indicates quantities computed at the time
$\tau_{*}$ when the non-attractor phase ends. Unlike the
attractor case, on super-horizon scales $\zeta$ is
not conserved: $\dot{\zeta}=3H\zeta$. From Eq.~(\ref{ratt}) one
can compute the power spectrum at the end of the non-attractor
($\tau=\tau_{*}$) era for modes that are already super-horizon
by that time ($|k\,c_{s}\tau_{*}|\ll 1$).  It is $P_{\zeta}(k)
=(4\,k^3)^{-1}(H_{*}/M_{P})^2(\epsilon_{*}c_{s})^{-1}$. The tensor power spectrum is likewise $P_\gamma(k) =
(1/k^3)\left(H_*/M_P\right)^2$.

\subsection{Tensor-scalar-scalar correlator}

Prior work considered the scalar-scalar-scalar three-point
function in $k$-inflation
\cite{Garriga:1999vw,Seery:2005wm,Chen:2006nt,Chen:2013aj,Chen:2013eea} as well as the scalar-tensor-tensor correlator \cite{Abolhasani:2013vaa}.
Here we calculate the squeezed limit of the primordial
tensor-scalar-scalar bispectrum in $k$-inflation and in
particular in the non-attractor model \cite{Chen:2013aj,Chen:2013eea}, focusing on modes that
left the horizon  before the beginning of the attractor phase\footnote{The tensor-scalar-scalar correlator has also been calculated in unpublished work by A. H. Tajdini and H. Firouzjahi.}.

\begin{figure}
\begin{center}
\includegraphics[scale=0.70]{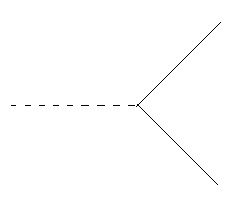}
\caption{Diagrammatic representation of the contribution in
     Eq.~(\ref{bisint}) to the tensor-scalar-scalar
     correlator. The dashed line is the graviton propagator,
     continuous lines are the scalars.} 
\label{fig2}
\end{center}
\end{figure}

The Schwinger-Keldysh (or \textsl{in-in}) formula
\cite{Schwinger:1960qe} (see also Ref.~\cite{Weinberg:2005vy} for a
detailed and comprehensive review of the formalism) is the
standard tool to calculate cosmological
correlation functions. In particular, the tree-level
contribution to the diagram depicted in Fig.~\ref{fig2} is,
\be\label{ia}
     \langle \gamma(\tau_{0})\zeta(\tau_{0})\zeta
     (\tau_{0}) \rangle =-i\,\int^{\tau_{0}}_{-\infty}
     d\tau\langle \left[\gamma(\tau_{0})\zeta(\tau_{0})
     \zeta(\tau_{0}),H_{\gamma\zeta^{2}}(\tau)\right]\rangle,
\ee
where $\left[\,,\right]$ denotes a quantum commutator and
$H_{\gamma\zeta^{2}}=-\,L_{\gamma\zeta^{2}}$ is the Hamiltonian
density (which can be obtained from Appendix~\ref{sec:tssaction})
to third order in fluctuations and $\tau_{0}$ is the
time of observation, so $\tau_{0}>\tau_{*}$. The time integral
in Eq.~(\ref{ia}) can be split into two integrals, respectively
for $\tau\in (-\infty,\tau_{*})$ and $\tau\in
(\tau_{*},\tau_{0})$. For the first integral we use
the non-attractor solution, Eq.~(\ref{ratt}), and take $\epsilon\sim
a^{\eta}$; for the second integral the attractor solution,
Eq.~(\ref{Att}) is employed, and $\epsilon$ is treated as a
constant and evaluated at $\tau_{*}$.

The Fourier-space field operators for the scalar and tensor
perturbations are
\begin{equation}
     \zeta_{\vec{k}}(t)\equiv
     a_{\vec{k}}\zeta_{k}(t)+a^{\dagger}_{-\vec{k}} \zeta^{*}_{k}(t),\quad
     \gamma_{ij,\vec{k}}(t)\equiv \sum_{\lambda=\pm}
     \epsilon_{ij}^{\lambda}(\hat{k})\left[b_{\vec{k}}^{\lambda}
     \gamma_{k}(t)+(b^{\lambda})^{\dagger}_{-\vec{k}}
     \gamma^{*}_{k}(t)\right],
\end{equation}
and the creation/annihilation operators for the scalar and the
tensor sectors obey the usual commutation relations.  The scalar
wavefunction was given in Eq.~(\ref{ratt}), and the tensor
wavefunction is the standard one,
\bea
     \gamma_{k}(\tau) & = & \frac{H_{*}}{M_{P}\sqrt{k^{3}}}
     \left(1+ik\tau\right)e^{-ik\tau},
\eea
where $d\tau=a\,dt$ is the conformal time and $``\,^{*}\,"$ as
usual indicates quantities evaluated at the end of the
non-attractor phase. 

We now evaluate Eq.~(\ref{ia}) to obtain the
tensor-scalar-scalar bispectrum.  Following
Ref.~\cite{Berezhiani:2013ewa}, we define a primed three-point
correlator $\VEV{\cdots}'$ by
\begin{equation}
     \VEV{\gamma^{\lambda}_{\vec{k}_{1}}\zeta_{\vec{k}_{1}} \zeta_{\vec{k}_{2}}
     } \equiv (2\pi)^{3}  \delta_D^{(3)}(\vec{k}_{1}+
     \vec{k}_{2}+\vec{k}_{3})
     \VEV{\gamma^{\lambda}_{\vec{k}_{1}}\zeta_{\vec{k}_{1}}\zeta_{\vec{k}_{2}}}'.
\end{equation}
We then define the tensor-scalar-scalar bispectrum
$\mathcal{B}(k_1,k_2,k_3)$ by
\begin{equation}
     \VEV{ \gamma^{\lambda}_{\vec{k}_{1}}\zeta_{\vec{k}_{2}}
     \zeta_{\vec{k}_{3}}}'\equiv \epsilon^{\lambda}_{ij}(\hat k_1) \hat
     k_{2i} \hat k_{3j} \mathcal{B}(k_{1},k_{2},k_{3}),
\end{equation}
where $\epsilon_{ij}^{\lambda}(\hat k_1)$ is the polarization tensor of the
tensor mode.  Connection with the functions $K_{ij}$ and
$A_{ij}$ in Section~\ref{sec:origin} can be made by identifying,
\begin{equation}
     k_{2i} k_{3j} \mathcal{B}(k_1,k_2,k_3) = \frac{1}{2}
     P_\gamma(k_1)(K_{ij}+A_{ij}).
\end{equation}

We first evaluate Eq.~(\ref{ia}),
with the interactions given in the first line of
Eq.~(\ref{intL}). We label this correlator
with a subscript ``\textsl{int}" to distinguish it
from the contributions to the total tensor-scalar-scalar
correlator arising from the field redefinitions (we further elaborate on these terms later in this Section and
report them in Appendix~\ref{sec:tssaction}),
\bea\label{bisint}
     \mathcal{B}(k_1,k_2,k_3)_{int}
     =\mathcal{B}_{[\gamma\partial\zeta\partial\zeta]}(k_{1},k_{2},k_{3})
     +\mathcal{B}_{[\partial^{2}\gamma\partial\chi\partial\chi]}
     (k_{1},k_{2},k_{3})+\mathcal{B}_{[\dot{\gamma}\partial\zeta\partial\chi]}
     (k_{1},k_{2},k_{3}).
\eea

The first interaction, $(\gamma\partial\zeta\partial\zeta)$,
is the standard attractor-phase result. This can be easily anticipated by
noticing that the integrand function for the non-attractor phase
part of the integral is equal to the integrand function of the
attractor period; indeed for the non-attractor phase the
non-standard time dependence of the wave functions for the
scalar fluctuations and the novel time dependence of the
slow-roll parameter cancel out to leave the typical expressions
that apply to the usual phase,
\be
     \epsilon \times \left(\zeta^{non-attr}\right)^{2} =
     \epsilon_{*}\frac{\tau^{6}}{\tau^{6}_{*}} \times
     \left(\zeta^{attr}\frac{\tau_{*}^{3}}{\tau^{3}}\right)^{2}
     =\epsilon_{*} \times \left(\zeta^{attr}\right)^{2}.
\ee
The result for this first contribution in the squeezed limit,
$k_{1}\ll k_{2}\sim k_{3}$, is
\be\label{fc}
     \mathcal{B}_{[\gamma\partial\zeta\partial\zeta]}(k_{L},k_{S},k_{S})
     =
     \frac{3}{8} \frac{H_{*}^{4}}{M_{P}^{4}}
     \left(\frac{1}{\epsilon_{*}\,c_{s}}\right)
     \left(\frac{1}{k_{S}^{3}k_{L}^{3}}\right), 
\ee
where we redefined the momenta as $k_{1}=k_{L}$ (long-wavelength
mode) and $k_{2}=k_{S}$ (short-wavelength mode). The result in
Eq.~(\ref{fc}) satisfies the consistency relation for
tensor-scalar-scalar correlators to order $\mathcal{O}(k_{L}^{0})$,
\begin{eqnarray}
\label{cR}
     \mathcal{B}_{c.r.}(k_{L},k_{S},k_{S})& \equiv&-\frac{1}{2}
     P_{\gamma}(k_{L})
     P_{\zeta}(k_{S})\frac{\partial
     \ln\,P_{\zeta}(k_{S})}{\partial\ln\,k_{S}}.
\end{eqnarray}
The remaining two interactions amount to,
\begin{equation}\label{nn}
     \mathcal{B}_{[\partial^{2}\gamma\partial\chi\partial\chi]}(k_{L},k_{S},k_{S})
     +\mathcal{B}_{[\dot{\gamma}\partial\zeta\partial\chi]}(k_{L}
     ,k_{S},k_{S})=\frac{297}{32} \frac{H_{*}^{4}}{M_{P}^{4}}
     \left(\frac{1}{\epsilon_{*}c_{s}^{}}\right)
     \left(\frac{1}{k_{L}^{3}k_{S}^{3}}\right)
     \frac{\epsilon_{*}}{(\tau_{*}c_{s}k_{S})^{6}}
     \left(\frac{k_{L}}{c_{s}k_{S}}\right)^{2}.
\end{equation}
Although this contribution is suppressed relative to that in
Eq.~(\ref{fc}) by a power of
$\epsilon_{*}$ and by a coefficient $(k_{L}/c_{s}k_{S})^{2}$, it
is also characterized by a factor
$(k_{S}c_{s}\tau_{*})^{-6}$. The latter, for modes that left the
horizon\footnote{Remember, throughout this work, we are after
modes $k_{S}$ and $k_{L}$ that are super-horizon by the time
the non-attractor era ends.} before $\tau_{*}$, provides
an enhancement.

This term then has the potential to provide an important
contribution to the squeezed limit of the observable $\<\gamma
\zeta \zeta \>$. It is quadratic in the soft momentum $k_L$
and it therefore escapes the familiar consistency condition as written in Eq.~(\ref{simple}).

In fact, already in the standard attractor-phase scenario, one
would typically have terms which scale just like Eq.~(\ref{nn}),
that is as $\left(k_L/k_S\right)^2$. The difference
now relies on the fact that there is a $\tau^{*}$ dependence: it
is the imprint of a past non-attractor phase. It shows how the
novel time dependence characterizing the wavefunction and the
slow-roll parameters during the initial inflationary stage can
counteract the slow-roll suppression.  

Eq.~(\ref{nn}) does not exhaust all
the contributions at second order in slow roll. The complete
calculation requires the next-to-leading order
slow-roll corrections of the first interaction term in
Eq.~(\ref{intL}). However, just as in the standard scenario,
Eq.~(\ref{nn}) represents a typical contribution to the
three-point correlator, and it is as such that we calculate its
contribution to the observables in Section
\ref{sec:observations}. As for field
redefinitions like those in Eqs.~(\ref{fr1}) and (\ref{fr2}),
they do certainly contribute but they also fall into the ``late
time'' argument given above.  They therefore do not provide a
source to directly probe the  $\tau_{*}$ dependence of the
three-point function.

The tensor-scalar-scalar bispectrum has contributions also from
the field redefinitions Eqs.~(\ref{fr1})--(\ref{fr2}). However, their contributions
are sub-leading compared to Eq.~(\ref{bisint}). For the sake of completeness, we report their expressions
in Appendix~\ref{sec:tssaction}.

Before continuing we note that the interaction terms
$\dot \gamma \p \zeta \p \chi\, , \, \p^2 \gamma \p \chi \p \chi
$, precisely those probing the model-dependent part of the
squeezed limit, would be, in the attractor phase, clearly suppressed. It is the time dependence of $\epsilon$ in the
non-attractor case that is able to counteract the
suppression. However, this counteraction may not go on forever,
in the sense that it cannot be that increasingly
$\epsilon$-suppressed terms that keep playing an 
important role in the result.  These considerations provide a
bound on the value of $k_s \tau_{*}$ simply springing from the
consistency of the slow-roll expansion.

We now comment on the slow-roll expansion and on the magnitude
of the contributions in Eqs.~(\ref{fc}) and (\ref{nn}). 
In slow-roll, given each interaction, its contributions to
higher-order correlators can be expanded in powers of the small
perturbative parameter $\epsilon$. As a simple consistency criterion
on this expansion, one requires that contributions given by
increasing orders in $\epsilon$ get smaller and that
the expansion converges. This is easily implemented during the
attractor phase, when $\epsilon$ is constant. The time dependence of
$\epsilon$ characterizing the non-attractor stage ought not to spoil
this feature: higher-order slow-roll corrections cannot become
leading.

There are three expansion parameters to keep in mind:
$\epsilon$, $k_{L}/(c_{s}k_{S})$ (we are probing the squeezed
limit), and $k_{S}c_{s}\tau_{*}$, which is smaller than unity for
modes that exited during the non-attractor phase. The result in
Eq.~(\ref{fc}), for instance, represents the leading-order
contribution in $k_{L}/(c_{s}k_{S})$, from the interaction
$\gamma\partial\zeta\partial\zeta$. Corrections to Eq.~(\ref{fc})
from this interaction include higher powers of
$k_{L}/(c_{s}k_{S})$,
\begin{equation}
     \mathcal{B}_{[\gamma\partial\zeta\partial\zeta]} \sim
     \frac{1}{k_{S}^{3}k_{L}^{3}}\left[1+\mathcal{O}
     \left(\frac{k_{L}}{c_{s}k_{S}}\right)^{n}\right],
     \quad\quad\quad (n>0),
\end{equation}
which one neglects in the squeezed limit. Similarly,
Eq.~(\ref{nn}) represents the leading-order contribution in
powers of $k_{L}/(c_{s}k_{S})$ for the other two interactions,
$\partial^{2}\gamma\partial\chi\partial\chi$ and
$\dot{\gamma}\partial\zeta\partial\chi$. Note that
Eq.~(\ref{nn}) is also the leading term in powers of
$k_{S}c_{s}\tau_{*}$,
\begin{equation}
     \mathcal{B}_{[\partial^{2}\gamma\partial\chi\partial\chi]
     +[\dot{\gamma}\partial\zeta\partial\chi]} \sim
     \frac{\epsilon_{*}}{k_{S}^{3}k_{L}^{3}}
     \left[\left(\frac{k_{L}}{c_{s}k_{S}}\right)^{2}+\mathcal{O}
     \left(\frac{k_{L}}{c_{s}k_{S}}\right)^{2+n}\right]
     \left[\frac{1}{q^{6}}+\mathcal{O}(q^{m})\right],\quad(n>0,\,\,m\geq
     0),
\end{equation}
where we defined $q\equiv(-k_{S}c_{s}\tau_{*})$. 
To enforce the criterion on typical-interaction terms at each
order in slow-roll we proceed as follows: we require the
$\gamma\p\zeta\p\zeta$ contribution to be larger than its
counterpart at higher powers of $\epsilon$.  We do the same for
$\p^2 \gamma\p\chi\p\chi, \, \dot\gamma\p\zeta\p\chi$. Also,
since the term in Eq.~(\ref{nn}) is $\epsilon_{*}$ suppressed with
respect to the one in Eq.~(\ref{fc}), we require the latter be
larger than the former. 

All these conditions are readily met by requiring,
\begin{equation}\label{condd1}
     q^{6}>\epsilon_{*},
\end{equation}
which is a reflection of the fact that a longer duration for the
non-attractor phase will inevitably lead to a more marked
imprint on the subsequent eras.  Eq.~(\ref{condd1}) leads then
to a (conservative) bound on the duration of such a stage. 
Notice that this condition arises naturally also from some
simple considerations. One denotes by $N_{*}$ the total number
of $e$-foldings, $e^{N_{*}}=(a_{*}/a_{in})$,
for the non-attractor phase, where $a_{*}=a(\tau_{*})$
($\tau_{*}$ being the time at which the non-attractor phase
ends) and $a_{in}$ is the value of the scale factor at the
beginning of the non-attractor phase. Considering modes that
exit the horizon before the end of the non-attractor phase, one
writes
\begin{equation}\label{ep1}
     -k_{S}c_{s}\tau_{*}= \frac{a_{S}H}{a_{*}H}
     =\frac{a_{S}}{a_{*}} >\frac{a_{in}}{a_{*}}=e^{-N_{*}},
\end{equation}
where $a_{S}$ is the value of the scale factor at the time the
mode $k_{S}$ leaves the horizon, and $c_{s}k_{S}=a_{S}H$. During
the non-attractor phase the parameter
$\epsilon\equiv-\dot{H}/H^{2}$ is a decreasing function of time,
$\epsilon(\tau)=\epsilon_{*} \left(\tau/\tau_{*}\right)^{6}$,
for $\tau<\tau_{*}$.
Requiring that $\epsilon$ is a small quantity for the whole
duration of the non-attractor phase (and of course later as
well), then $\epsilon_{*}< e^{-6N_{*}}$,
which is consistent with Eqs.~(\ref{ep1}) and (\ref{condd1}). 

In and of itself, the previous equation does not provide a strong
constraint on the parameter space of the model under
scrutiny. However, if we take a tensor-to-scalar ration $r= 16\,
\epsilon_{*} c_s \sim 0.1$ and assume $c_s=1$, then we find a
non-attractor phase that lasts about one $e$-fold. A longer
non-standard stage requires a smaller $\epsilon_{*}$.

To get a grasp of how the relative size of the various
interaction contributions translate into bounds one may, for
example, require that the contribution in Eq.~(\ref{nn}) be smaller
but within the same order of magnitude of Eq.~(\ref{fc}). This would
result in a ratio $k_L/k_S$ which is still small, of order $1/10$.

\section{Solid Inflation}
\label{sec:solid}

\subsection{The Model}
Recently, an intriguing inflationary model has been put forward
in Ref.~\cite{Endlich:2012pz}. Although the symmetry-breaking
pattern of this theory is far from that in the standard
picture, it nevertheless results in a well-controlled
inflationary phase that produces signatures quite distinct from
those of the standard scenario. Of interest here is the fact
that this model entails no adiabatic modes during the entirety
of the inflationary era.  We begin by reprising the theory and
then move on to discuss a consistency-condition--violating
scalar-scalar-tensor three-point function. For a more thorough
treatment we refer the reader to the original work in
Ref.~\cite{Endlich:2012pz} and follow-ups in
Refs.~\cite{Endlich:2013dma,Nicolis:2013lma,Endlich:2013jia} (for more on SI see also \cite{Bartolo:2013msa,Akhshik:2014gja}).

Solid inflation describes an homogeneous and isotropic cosmological background
even though it involves a scalar-field background that is
space-dependent and breaks spatial-translation and rotational
invariance. The key point is to require some ``internal"
symmetries in field space so as to make the combined
spatial$+$internal transformations a symmetry of the theory.

The symmetry-breaking background is $\langle \phi^{I} \rangle=
x^{I}$, 
but the theory is also endowed with internal symmetries under,
\bea
     \phi^{I}\rightarrow \phi^{I} + A^{I};\qquad
     \phi^{I}\rightarrow O^{I}_{J} \phi^J, \quad (O^{I}_{J} \in
     SO(3)). \label{symm}
\eea
One can think of the $\phi^{I}$'s as internal comoving
coordinates around which, once time dependence is accounted for,
describe a solid and its volume-element
position, $\bar x=\bar{x}(t, \phi^{I})$. But at each $t$ this
might be inverted to give $\phi^{I}=\phi^{I}(t,\bar{x})$. The
latter form turns out to be more convenient, as one handles
spacetime symmetries in the more familiar fashion.  The 
treatment then reduces, in flat space,\footnote{The extension to
cosmological solutions with dynamical gravity is not
particularly involved, but we omit details here.} to that of
a relativistic low-energy effective theory of the three Poincare
scalars $\phi^{I}$ endowed with a number of symmetries (those
dictated by the request of homogeneity and isotropy in the
background).

These simple considerations coupled with Eq.~(\ref{symm}) are
enough to greatly constrain the form of the action at lowest
order in the derivative expansion,
\bea
     S=\int d^4 x\, F\left(X, Y,Z\right)\quad \equiv \quad  \int
     d^4 x \,F\left([B],\,\,
     \frac{[B^2]}{[B]^2},\,\,\frac{[B^3]}{[B]^3}   \right), 
\eea
where $B^{IJ}= (\p_{\mu}\phi^{I})( \p^{\mu}\phi^{J})$ and the first
invariant (under Eq.~(\ref{symm})) variable, $X\equiv [B]\equiv
Tr[B^{IJ}]$,  has been chosen to keep track of the overall size
of the system.  The background will spontaneously break some of
the symmetries, a breaking to which Goldstone bosons will be
associated.\footnote{The logic here is the same as in the
effective-field-theory approach but the
symmetry-breaking pattern is different and this has crucial
consequences, some of which we shall report below.} These are
fluctuations $\phi^{I}=\phi^{I}(x^{I}+ \pi^{I})$ around the
background solution $\phi^{I}(x^{I})$.

\subsection{Perturbations}

Expanding in the flat-space action one gets, at quadratic order
\cite{Endlich:2012pz}, 
\begin{equation}
     S^{(2)} = \int d^4 x \left[-\frac{1}{3}  F_X X
     \dot{\vec \pi}^2 +\big( \frac{1}{3} F_X X +\frac{6}{27}
     (F_Y+F_Z) \big)  (\p_i \pi_j )^2   + \big( \frac{1}{9}
     F_{XX} X^2 +\frac{2}{27} (F_Y+F_Z) \big)  (\vec \nabla
     \cdot \vec \pi )^2\right] ,\nn\\
\label{quad}
\end{equation}
where the derivatives $F_{(.)}$ are calculated on the
background. These Goldstone modes are phonons of the solid, and
they can be split into longitudinal and transverse
 components $\bar{\pi}_L$ and $\bar{\pi}_{T}$ with
 associated speeds of propagation \cite{Endlich:2012pz},
\bea \label{cLcT}
     c_L^2 = 1 +  \frac{2}{3} \frac{F_{XX} X^2}{F_X X} +
     \frac{8}{9} \frac{(F_Y + F_Z)}{F_X X} \; , \qquad c_T^2 = 1
     + \frac{2}{3} \frac{(F_Y + F_Z)}{F_X X}.
\label{speed}
\eea
Gravity can be turned on by promoting $\eta_{\mu
\nu}$ to a more general $g_{\mu \nu}$, introducing the
corresponding measure and minimally coupling gravity to the
$\phi^{I}$ fields. The resulting stress-energy tensor is indeed
homogeneous and isotropic as desired. To sustain a
superluminality-free slow-roll phase without
strong-coupling issues, the parameters must satisfy
\cite{Endlich:2012pz},
 \bea
      \ep= \frac{3}{a^2} \frac{F_X}{F}=\frac{\p {\rm log} F}{\p
      {\rm log} X}\ll 1\quad(s.r.); \qquad 0<F_Y+F_Z <
      \frac{3}{8}\epsilon |F|\quad ({\rm
      luminality}). \label{conditions1}
 \eea
The smallness of $F_X$ can be linked with the breaking of a
scaling symmetry (under which both $Y,Z$ are already
invariant, as one might expect, $X$ being the only
variable that carries the overall size information of the
system) and therefore be stable under quantum corrections.  This
is not true, though, for the latter condition in
Eq.~(\ref{conditions1}) which is at this stage an assumption
that might require some tuning.   The validity of perturbation
theory requires that the condition,
\bea
     \ep\, c_L^3\gg (H/M_{P})^{2/3}\quad({\rm pert. theory}),  
\label{sc}
\eea
holds.  The smallness of the slow-roll parameter $s=
\dot{c}_s/(H c_s)$ is automatic if the above conditions are met
and therefore generates no more bounds. 

One of them is the fact that the ``clock'' regulating a beginning
and signaling the end for inflation is entirely due to the
metric field. One might choose a number of gauge-invariant
observables as the physical clock, such as the energy density
and pressure, or $X$ itself, but the unperturbed $\phi^{I}$s by
themselves have no time dependence. This is in stark contrast
with the physical-clock role played by the inflaton in most
inflationary mechanisms. It is a fact of consequence because
some of the predictions of the model will depend on which
variable is used to signal the end of inflation and trigger
reheating.

The Goldstone-boson picture has been already employed in the
literature to describe inflation, most notably in the
effective-field-theory approach to inflation
\cite{Cheung:2007st}. In  this approach, the symmetry-breaking
pattern is the more familiar one where it is time
diffeomorphisms that are broken by the Goldstone; one can
show\footnote{These differences appear in many facets;
e.g., in solid inflation (SI) the physical clock is entirely due
to the metric, in SI there are no adiabatic fluctuations, etc..}
that the two scenarios can\textit{not} be mapped into each other. 

In solid inflation  tensor and scalar gauge-invariant fluctuations
are not adiabatic and are not conserved outside the
horizon. This is, as we have seen, the scenario where violations
of consistency conditions take place. The physical picture as to
why this happens is \cite{Endlich:2012pz} that adiabatic mode
can be reabsorbed by a time shift of the background if the
wavelength of the fluctuation is large enough.  However, solid
can experience anisotropic stresses, non-scalar perturbations,
that cannot be re-absorbed as a time shift.

With these considerations it is not surprising that the
solutions to the equations of motion that emerge from the
following quadratic scalar, vector and tensor Lagrangians are
not constant outside the horizon:
\begin{eqnarray}
     S^{(2)} & =&  S^{(2)}_\gamma+S^{(2)}_{T}+S^{(2)}_{L} \\
     \label{graviton quadratic action}
     S^{(2)}_\gamma &=& \frac14 {M_P^2} \int dt\,d^3x
     \,a^3\Big[\frac12 \dot{\gamma}_{ij}^2 -\frac{1}{2 a^2}
     \big(\partial_m \gamma_{ij}\big)^2 +2\dot{H}c_T^2 \,
     \gamma_{ij}^2 \Big]\\
     \label{transverse quadratic action}
     S^{(2)}_{T} &=&M_P^2 \int dt \int_{\vec k} \,a^3
     \bigg[\frac{ k^2/4}{1-k^2/4a^2\dot{H}} \, \big|
     \dot{\pi}_T^i \big|^2 +\dot{H}c_T^2 \, k^2 \big|\pi_T^i
     \big|^2 \bigg]\\ 
     \label{longitudinal quadratic action}
     S^{(2)}_{L} &= &M_P^2 \int dt \int_{\vec k}  \, a^3 \bigg[
     \frac{ k^2/3}{1-k^2/3a^2\dot{H}}\big|\dot{\pi}_L
     -({\dot{H}}/{H})\pi_L\big|^2+\dot{H}c_L^2 \, k^2 \big|
     \pi_L  \big|^2 \bigg].
\end{eqnarray}

The mode functions for tensor fluctuations to first order in
slow-roll are,
\begin{eqnarray}
     \gamma_{k}(\tau)&=&
     (-\tau)^{3/2+\epsilon_c} \frac{H_c(1-\epsilon_c)}{M_P}
     \sqrt{\frac{\pi}{2}}\,(-\tau_c)^{-\epsilon_c}
     e^{\frac{i\pi}{2}(\nu_T +\frac{1}{2})} \,
     H_{\nu_T}^{(1)}(-k\tau) \;, \,\,\,  \nu_T \simeq
     \frac{3}{2}+\epsilon_c -\frac{4}{3} c_{T,c}^2
     \epsilon_c\,,\nonumber\\
\end{eqnarray}
where $c_{T}$ and $c_{L}$ are the transvere and longitudinal
propagation speed, $H^{(1)}$  is a Hankel function, and the
subscript ``c'' indicates quantities evaluated at some reference
time $\tau_{c}$, chosen here as the horizon exit time of the
longest modes relevant for observations.

For curvature fluctuations, the mode functions to first order in
slow-roll are more involved so we report for simplicity their
super-horizon limit ($-k\tau\rightarrow 0$),
\begin{equation}
     \zeta_{k}(\tau) =
     \left(\frac{\tau}{\tau_c}\right)^{\frac{4}{3}c_{T,c}^2\epsilon_c}(-
     c_{L,c}k \tau_c)^{c_{L,c}^2\epsilon_c-5s_c/2-\eta_c/2}
     \left(\frac{H_c}{\sqrt{4\epsilon_c}M_P c_{L,c}^{5/2}\,k^{3/2}}
     +\mathcal{O}(\ep^{1/2})\right),
\end{equation}
where $s\equiv\dot{c}_{L}/H\,c_{L}$ . See
Ref.~\cite{Endlich:2012pz} for further details.

Once the wave functions are obtained, it is straightforward to
calculate the power spectra\footnote{In SI the gauge-invariant
variables $\mathcal{R}$ and $\zeta$ are not equal, not even at
late times. They will be after reheating but one needs to choose
which quantities to focus on: $\langle\zeta\zeta...\rangle$ or
$\langle \mathcal{R}\mathcal{R}... \rangle$. The choice is
decided by the fact that in this model
$\langle\zeta\zeta\rangle$ must be continuous at
reheating.} at late times for scalar and tensor modes, they have
the following expressions:
\begin{equation}
     P_{\zeta}(k)=\frac{H_c^2}{4 \epsilon_c c_{L,c}^5 M_P^2}
     \frac{1}{k^3}
     \frac{(\tau/\tau_c)^{8c_{L,c}^2\ep_c/3}}{(-c_{L,c}k
     \tau_c)^{5s_c-2c_{L,c}^2\epsilon_c+\eta_c}}  ,\quad
     P_{\gamma}(k) = \frac{H_c^2}{M_{P}^2}  \frac{1}{k^3}
     \frac{(\tau/\tau_c)^{8c_{T,c}^2\ep_c/3}}{(-k\tau_c)^{-2c_{L,c}^2\ep_c}}.
\end{equation}
The spectral indexes for scalars and tensors,
\begin{equation}
     n_S-1 \simeq 2\epsilon_c c_{L,c}^2-5s_c-\eta_c , \qquad
     n_{T}\simeq 2\,c_{L,c}^{2}\epsilon_{c},
\end{equation}
can be read directly from the expressions above.  Note that the
tensor tilt is blue. The tensor-to-scalar ratio then has the
form:
\begin{equation}
     \label{scalar to tensor ratio}
     r \sim \epsilon \, c_L^5 \; .
\end{equation}
For the estimates in Section \ref{sec:observations}, we use the slow-roll
($\epsilon_c \to 0$) approximations,
\begin{equation}
     P_{\zeta}(k)=\frac{H_c^2}{4 \epsilon_c c_{L,c}^5 M_P^2} 
     \frac{1}{k^3} ,\qquad
     P_{\gamma}(k) = \frac{H_c^2}{M_{P}^2} \, \frac{1}{k^3},
\end{equation}
for these power spectra.

\subsection{Violations of the consistency conditions}

The squeezed limit of the scalar bispectrum is,
\begin{eqnarray}
     \mathcal{B}_{\zeta\zeta\zeta}(k_{L},k_{S},k_{S})=-\frac{20
     }{9}\frac{F_Y}{F} \frac{1}{ c_L^2\epsilon} \,\left(1-3
     \cos^2 \theta \right)\, P_{\zeta}(k_{L})P_{\zeta}(k_{S}),
     \label{sq2}
\end{eqnarray}
where $\theta$ is the angle between $\vec{k}_{L}$ and
$\vec{k}_{S}$. Eq.~(\ref{sq2}) manifestly violates the
consistency conditions. The shape of non-Gaussianity for solid
inflation has a very small overlap with the local template.
However if one, for lack of a better option, relies on Planck
$f_{NL}^{local}$ findings in order to constrain the parameters
in Eq.~(\ref{sq2}), then it is safer to assume $F_{Y}\ll \,F$.

The tensor-scalar-scalar bispectrum reads \cite{Endlich:2013jia},
\begin{equation}\label{ccss}
     \mathcal{B}_{\gamma\zeta\zeta}(k_{L},k_{S},k_{S})=
     -\frac{10}{9} \frac{F_Y}{F}\,\frac{1}{c_L^2
     \epsilon}P_{\gamma}(k_{L})P_{\zeta}(k_{S})
     \simeq -\frac{5}{18}
     \frac{F_Y}{F}\,\frac{1}{c_L^7
     \epsilon^{3}}\frac{H^{4}}{M_{P}^{4}}\frac{1}{k_{L}^{3}k_{S}^{3}}.
\end{equation}
Notice that, unless $(F_Y/F)(c_L^2 \epsilon)^{-1}=-27/20$,
Eq.~(\ref{ccss}) violates the consistency condition, Eq.~(\ref{cR}).
The bounds to be aware of at this stage are the one in
Eq.~(\ref{sc}) and the ones resulting from the luminality
condition on both $c_T$ and $c_L$ , which are related by
$c_T^2=(3/4)\left[1+c_L^2-(2/3)\epsilon +(1/3)\eta\right]$,
to all orders in $\epsilon$ and $\eta$.

\section{Observational signatures}
\label{sec:observations}

Primordial scalar perturbations give rise to temperature
fluctuations in the CMB and to mass-density perturbations in the
late Universe. These late-time mass-density perturbations can be
mapped with some precision through their effects on the galaxy
distribution, once the effects of galaxy-bias are taken into
account, and they can also be mapped through weak gravitational
lensing.  The effects of tensor metric perturbations can be
observed through measurements of CMB fluctuations, and in
particular, through measurements of the B mode of the CMB
polarization.  The effects of tensor metric perturbations may
also some day be seen in direct gravitational-wave searches
\cite{Liddle:1993zj,BarKana:1994bu,Turner:1996ck,Smith:2005mm,Chongchitnan:2006pe,Smith:2008pf,Kuroyanagi:2014qaa,Jinno:2014qka},
but these observations will map only short-wavelength tensor
modes.  Some of the lensing/CMB/large-scale-structure
observations discussed in the Introduction may some day map
larger-scale tensor modes, but those measurements are some way
in the future.  Even these measurements will probably not
forward to directly map the three-dimensional primordial
tensor-perturbation field at the largest scales that we will
encounter shortly.

The correlations of primordial tensor perturbations with primordial scalar 
perturbations can, however, have observational
consequences for the mass distribution, even if the tensor
perturbation cannot be detected directly.  
In the absence of a
tensor-scalar-scalar bispectrum (and in the approximation that
the scalar-scalar-scalar bispectrum is small), the primordial
scalar perturbation is Gaussian and statistically isotropic.
The tensor-scalar-scalar bispectrum will, however, induce
an apparent local departure from statistical isotropy
\cite{Giddings:2011zd,Dai:2013ikl,Dai:2013kra,Schmidt:2013gwa} and a
characteristic non-Gaussian four-point function
\cite{Seery:2008ax,Jeong:2012df,Brahma:2013rua} in primordial
perturbations.  These effects can be sought in the CMB and in
large-scale structure.  The depature from statistical isotropy
arises primarily from gravitational waves of wavelengths larger
than the galaxy-survey size, while the non-Gaussian effects may
arise from gravitational waves of wavelengths comparable to the
survey size.  Departures from statistical isotropy are therefore
expected to be most significant, relative to smaller-scale
non-Gaussianity, for models where the bispectrum peaks
dramatically in the squeezed limit.  As we will see, this is
what happens for non-attractor inflation, and so we will not
work out the smaller-scale non-Gaussianity expected in this
model.  The k-dependence of the tensor-scalar-scalar bispectrum for solid inflation
is, as we will see, closer to the behavior familiar from SFSR, and so we will
work out constraints and forecasts for the observability of the
small-scale non-Gaussian effects induced in the mass
distribution by solid inflation.

\subsection{Local Power Quadrupole}

Here we calculate the local power quadrupole, the observable
that follows from the squeezed limit of the tensor-scalar-scalar
bispectrum.  We begin by summarizing the main results derived
above.  

We focus in this work on the observable consequences of
this bispectrum in the squeezed limit, $k_L \ll k_S$, where $k_L
\equiv k_1$ and $k_S \equiv k_2 \simeq k_3$. In both models we first account for a contribution of the form given in Eq.~(\ref{cR})
in the squeezed limit. This is the contribution that arises
from the consistency condition between the squeezed-limit
bispectrum and the scalar and tensor power spectra $P_\gamma(k_L)$
and $P_\zeta(k_S)$, respectively.  This part of the bispectrum
gives rise to an infrared-divergent contribution to the local
power quadrupole moment that is then cancelled by a similarly
infrared-divergent late-time effect
\cite{Pajer:2013ana,Dai:2013kra} leaving a small, but nonzero
and observable (at least in principle), local power quadrupole
\cite{Dai:2013kra,Schmidt:2013gwa}.

What we are interested in here, though, is the part of the
bispectrum that violates\footnote{As mentioned, for non-attractor inflation the violation is to be understood in the sense of Subsection \ref{preview}.} the consistency condition.  For
non-attractor inflation, this was found to be [cf., Eq.~(\ref{nn})],
\begin{eqnarray}
\label{naB}
     \mathcal{B}_{\rm na, {\slashed{\rm cc} } }(k_{L},k_{S},k_{S})
     =\frac{297}{32}\frac{H_{*}^{4}}{M_{P}^{4}}
     \left(\frac{1}{\epsilon_{*}c_{s}^{}}\right)
     \left(\frac{1}{k_{L}^{3}k_{S}^{3}}\right)\epsilon_{*}
     \frac{1}{(\tau_{*}c_{s}k_{S})^{6}}
     \left(\frac{k_{L}}{c_{s}k_{S}}\right)^{2}.
\end{eqnarray}
For solid inflation it is, from Eq.~(\ref{ccss}),
\begin{eqnarray}
     \mathcal{B}_{\rm si,
	  {\slashed{\rm cc}}}(k_{L},k_{S},k_{S}) =
     -\frac{5}{18} \left(\frac{F_Y}{F} \frac{1}{c_L^2 \epsilon}
     + \frac{27}{20} \right)
     \frac{1}{c_L^5
     \epsilon^2}\frac{H^{4}}{M_{P}^{4}}\frac{1}{k_{L}^{3}k_{S}^{3}}\,.
\end{eqnarray}

The existence of gravitational waves with wavelengths long
compared with the distances over which observations are
performed (e.g., for the CMB, our observable horizon) gives rise
to an apparent local departure from statistical isotropy.  In
other words, the rms amplitudes of Fourier modes of the same
wavenumber but different directions may differ.  An individual
Fourier mode $\gamma_p(\vec k_L)$ of the tensor field\footnote{Any collection of modes with wavelengths much longer than the horizon will be undistinguishable from one another within any single horizon. One effectively has a single, long-wavelength, mode.} gives rise
to a local matter, or curvature, power spectrum,
\begin{equation}
     P_\zeta(\vec k_S) |_{\gamma_p(\vec k_L)} = P_\zeta(k_S) \left
     [1 + \mathcal{Q}^p_{ij}(\vec k_L) \hat k_S^i \hat k_S^j
     \right],
\end{equation}
with power quadrupole,
\begin{equation}
     \mathcal{Q}^p_{ij}(\vec k_L) =
     \frac{\mathcal{B}_{\slashed{\rm
     cc}}(k_L,k_S,k_S)}{P_\gamma(k_L) P_\zeta(k_S)} \gamma^p_{ij}(\vec k_L),
\end{equation}
where $\mathcal{B}_{\slashed{\rm cc}}(k_L,k_S,k_S)$ is the
consistency-condition--violating part of the
tensor-scalar-scalar bispectrum.  The observed power quadrupole
is then obtained by summing over both gravitational-wave
polarizations $p=\{+,\times\}$ and Fourier wavevectors $\vec
k_L$.

The theory then predicts that this locally observed power
quadrupole has variance,
\begin{equation}
     \overline{\mathcal{Q}^2} \equiv \frac{8\pi}{15} \VEV{
     \mathcal{Q}_{ij} \mathcal Q^{ij}} = \frac{16}{15\pi}
     \int_{k_L^{\rm min}}^{k_S^{\rm min}} \, k_L^2 \, dk_L\, \left[
     \frac{\mathcal{B}_{\slashed{\rm
     cc}}(k_L,k_S,k_S)}{P_\gamma(k_L) P_\zeta(k_S)} \right]^2 P_\gamma(k_L).
\end{equation}
Here, the upper limit of integration, $k_S^{\rm min}$, is the
smallest wavenumber probed by the observations.  The lower
limit, $k_L^{\rm min}$, corresponds to the longest-wavelength
gravitational-wave mode produced during inflation.

Using $P_\gamma(k) = (1/k^3)(H_*/M_P)^2$, the result for
non-attractor inflation is
\begin{equation}
     \overline{\mathcal{Q}_{\rm na}^2} = \frac{64}{15\pi} \left(
     \frac{297}{32}\right)^2 \left(\frac{H_*}{M_{P}}\right)^2
     \left(\frac{k_S^{\rm min}}{c_S k_S} \right)^4 \frac{
     \epsilon_*^2}{(c_s k_S \tau_*)^{12}}.
\end{equation}
The $k_L^2$ dependence of $\mathcal{B}/P_\gamma$ assures that
the result does not depend on the infrared cutoff $k_L^{\rm
min}$.   The falloff of the power quadrupole with increasing
$k_S$ is so steep that the observability
of the signal will depend almost entirely on the sensitivity to
a power quadrupole on the very largest scales.  Roughly
speaking, the observational upper limit will be
$\overline{\mathcal{Q}_{na}^2} \lesssim 1$ for $k_S\sim H_0$,
the Hubble parameter today.  This translates
into a bound $\tau_* \gtrsim H_0^{-1}$, which, given the very
strong dependence of the quadrupole on $k_S$, is fairly
insensitive to other model parameters.  We thus infer that the
absence of any grotesque departure of the components of the CMB
quadrupole from statistical isotropy tells us that a
non-attractor phase of inflation must have ended no later than
the time that our current Universe exited the horizon during
inflation.

For solid inflation, the result is
\begin{equation}
     \overline{\mathcal{Q}_{\rm si}^2} = \frac{64} {15\pi}
     \left[
     \frac{5}{9} \left( \frac{F_Y}{F c_L^2\epsilon} +
     \frac{27}{20} \right) \right]^2
     \frac{H_c^2 }{\epsilon^2 M_{P}^2} \ln\left(
     \frac{k_S^{\rm min}}{k_L^{\rm min}}\right) \equiv A \ln\left(
     \frac{k_S^{\rm min}}{k_L^{\rm min}}\right),
\label{eqn:solidresult}
\end{equation}
which defines the prefactor $A$.  In this case, the power quadrupole
diverges logarithmically as $k_L^{\rm min} \to 0$, implying
sensitivity of the observable to very-long-wavelength modes,
something that does not arise for the part of the
bispectrum that satisfies the consistency condition.  The
observation that $\overline{\mathcal{Q}_{\rm na}^2} \lesssim 1$ for
$k_S\sim H_0$ then implies an upper limit $A
\left|\ln\left(k_L^{\rm min}H_0^{-1} \right) \right|\lesssim1$.

\subsection{Clustering fossils}

Here we consider the the characteristic non-Gaussian four-point
correlations in the scalar perturbation induced by coupling to
tensor modes.  Ref.~\cite{Jeong:2012df} provides a recipe for
measuring these correlations with a galaxy survey (or other
tracer of the three-dimensional mass distribution) and estimates
the detectability of the signal for single-field slow-roll
inflation.  
We first consider the case with the primordial bispectrum obeying the 
SFSR consistency relation including the late-time effects of tensor-scalar
coupling and projection effect, then move on to the case with solid inflation
and calculate the signal-to-noise ratio for measuring 
the signature of clustering fossils.

\subsubsection{Clustering fossils with the consistency relation}

The consistency relation dictates that the tensor-scalar-scalar
bispectrum takes the form in Eq.~(\ref{cR}) in the
squeezed limit.  This bispectrum
implies that in the presence of a Fourier mode $\gamma_p(\vec
k_L)$ of the tensor perturbation, the correlation between two
scalar-perturbation modes $\zeta(\vec k_1)$ and $\zeta(\vec
k_2)$ is \cite{Jeong:2012df},
\be
\left.
\left<
\zeta(\vec k_1)\zeta(\vec k_2)
\right>
\right|_{\gamma_p(\vec k_L)}
=
\delta^D_{\vec k_1+\vec k_2} P_\zeta(k_1)
-\frac12 \delta^D_{\vec k_L+\vec k_1+\vec k_2}
\frac{{\rm d} \ln P_\zeta(k_1)}{{\rm d}\ln k_1} 
P_\zeta(k_1)
\gamma_p(\vec k_L)
\hat{\varepsilon}_{ij}(\hat{k}_L)\hat{k}_1^i\hat{k}_2^j,
\label{eq:zeta_zeta_initial}
\ee
where we use the shorthand $\delta^D_{\vec k}\equiv (2\pi)^3
\delta_D(\vec k)$.
There is now a new term, in addition to the usual power spectrum, 
that correlates different Fourier modes of the scalar
perturbation.  This off-diagonal correlator can be understood as
a local rescaling, $k^2 \to (\delta_{ij} + \gamma_{ij}(\vec
k_K))k^ik^j$, of the wavevector.

Until inflation ends, all relevant density modes are outside the 
horizon with the local correlation function frozen with the form of 
Eq.~(\ref{eq:zeta_zeta_initial}). After inflation ends, density modes 
continuously come inside the horizon and evolve under the 
influence of the long-wavelength tensor field.  This yields a local 
density contrast,
\be
\left.
\delta(\vec k_S)\right|_{\gamma(\vec k_L)} = 2T_\delta(k_S) 
\left[
1-
\left(
\frac12 \frac{{\rm d}\ln T_{\delta}(k_S)}{{\rm d}\ln k_S}
+
S_N(k_L) 
\right)
\gamma^{ij}(\vec k_{L,i}) \hat{k}_{S,i}\hat{k}_{S,j} 
\right]\zeta(\vec k_S),
\ee 
where $T_\delta(k_S)$ and $T_\gamma(k_L)$ are the transfer
functions for, respectively, the density field and gravitational
wave, and
\be
     S_N(k_L) \simeq
     \frac35\left[1-\exp\left(-\frac{5}{42}k_L^2\eta^2\right)\right],
\ee
encodes the dynamical influence of the long-wavelength tensor mode 
$\gamma_p(\vec k_L)$ on the evolution of the small-scale scalar mode 
\cite{Dai:2013kra}.  At large scales ($k_L\to0$), the function
$S_N(k_L)$ vanishes; this obeys the causality demand that there
is no influence from super-horizon tensor modes on the evolution
of subhorizon scalar modes.  On small scales ($k_L\gg H_0$), the
function $S_N(k_L)$ asymptotes to $3/5$.  This then partially
cancels the primordial off-diagonal correlation in
Eq.~(\ref{eq:zeta_zeta_initial}) to yield an observed density
field that satisfies,
\be
\left.\left<
\delta(\vec k_1)
\delta(\vec k_2)\right>\right|_{\gamma_p(\vec k_L)}
\simeq 
\delta^D_{\vec k_1+\vec k_2} P_{\delta}(k_1)
-
\delta^D_{\vec k_1+\vec k_2 +\vec k_L}
\left[
\frac12
\frac{{\rm d}\ln P_\delta(k)}{{\rm d}\ln k}
+
2S_N(k_L) 
\right]
P_\delta(k_1)
\gamma^{ij}(\vec k_L) \hat{k}_{1}^i\hat{k}_{2}^j.
\ee

Finally, with the linear-bias parameter $b_g$, the observed
galaxy density  contrast is given in terms of the intrinsic
matter density contrast and the projection at the location of
galaxies and the line-of-sight as 
\cite{Jeong:2011as,Jeong:2012nu,Schmidt:2012nw,Dai:2013kra}:
\be
\delta^{\rm obs}_g
=
b_g
\left[
\delta
+
\Delta x^i\partial_i \delta
+
\left(b_e H \Delta t + \partial_i \Delta x^i\right)\delta
\right]
\simeq
b_g \delta 
-
\frac12 b_g T_\gamma
\gamma^{ij}x_j \partial_i \delta
\ee
with the temporal and spatial displacement ($\Delta t$ and $\Delta x^i$) 
of galaxies due to light deflection.
The dominant projection effect comes from large scales $k_L\ll
H_0$ where $\Delta t$ vanishes and $\Delta x^i =
-T_\gamma\gamma^{ij}x_j/2$ \cite{Schmidt:2012nw}.  Including
projection effects \cite{Dai:2013kra}, the observed
galaxy-density field thus satisfies, 
\begin{eqnarray}
\left.\left<
\delta_g(\vec k_1)
\delta_g(\vec k_2)\right>\right|_{\gamma_p(\vec k_L)}
&\simeq &
\delta^D_{\vec k_1+\vec k_2} P_g(k_1) \nonumber \\
& & -  \delta^D_{\vec k_1+\vec k_2 +\vec k_L}
\left[
\frac12 (1-T_\gamma)
\frac{{\rm d}\ln P_\delta(k_1)}{{\rm d}\ln k_1}
+
2 S_N(k_L) 
\right]
P_g(k_1) 
\gamma^{ij}(\vec k_L) \hat{k}_{1}^i\hat{k}_{2}^j .\nonumber \\
\label{eqn:correctedfossil}
\end{eqnarray}
This equation corrects the clustering-fossil result in
Ref.~\cite{Jeong:2012df} to take into account the cancellation
of the infrared divergence from the initial bispectrum in these
observables from projection effects
\cite{Pajer:2013ana,Dai:2013kra} leaving an observable ${\cal
O}(k_L^2)$ clustering fossil \cite{Dai:2013kra}.  

The inclusion here of late-time effects revises the result in
Fig.~2 of Ref.~\cite{Jeong:2012df} for bispectra that satisfy
the consistency condition.  After taking these effects into
account, the $3\sigma$ detection limits in Eq.~(10) and Fig.~2 of
Ref.~\cite{Jeong:2012df} are increased by a factor,
\be
\left(
\frac12 
\frac{{\rm d}\ln P_\delta(k_1)}{{\rm d}\ln k_1}
\right)^2
\left[
\frac12 (1-T_\gamma)
\frac{{\rm d}\ln P_\delta(k_1)}{{\rm d}\ln k_1}
+
2S_N(K) 
\right]^{-2}
\simeq  25
\ee
for $K \gtrsim k_H$.

\subsubsection{Clustering fossils in solid inflation}

The tensor-scalar-scalar correlator can be used to estimate the amplitude of tensor modes. One can construct an optimal variance estimator \cite{Jeong:2012df} for the tensor power spectrum, with variance $\sigma_{\gamma}$ given by
\bea
\label{var}
\sigma^{-2}_{\gamma}=\frac{1}{2}\sum_{\vec{k}_{L},p}\left[ k_{L}^{3} P_{p}^{n}(k_{L})\right]^{-2} \,\, ,
\eea
where $P_{p}^{n}$ is the noise power spectrum, defined as 
\bea
P_{p}^{n}(k_{L})=\left[\sum_{\vec{k}_{S}}\frac{|\mathcal{B}_{
	  {\slashed{\rm cc}}}(k_{L},k_{S},|\vec{k}_{L}-\vec{k}_{S}|)\epsilon_{ij}^{p}\hat{k}_{S}^{i}\hat{k}_{LS}^{j}|^{2}}{2 V P_{\gamma}^{2}(k_{L})P_{\zeta}^{tot}(k_{S})P_{\zeta}^{tot}(|\vec{k}_{L}-\vec{k}_{S}|)}\right]^{-1} \, \, .
\eea
Above $\hat{k}_{LS}\equiv (\vec{k}_{L}-\vec{k}_{S})/|\vec{k}_{L}-\vec{k}_{S}|$, $P_{\zeta}^{tot}$ is the total measured scalar power spectrum (i.e. including signal and noise) and $V$ is the total volume of the survey. Notice that the variance in Eq.~(\ref{var}) is inversely proportional to the variance of the quadrupole. As a result, for a given survey size, the larger the amplitude of the quadrupole, the larger the minimum amplitude of tensor modes that one is able to probe.

Given the similarity of the $k_L$ and $k_S$
dependences of scalar and tensor power spectra and
tensor-scalar-scalar bispectrum in solid inflation with those of
SFSR inflation
\begin{equation}
     \mathcal{B}_{\rm si,
	  {\slashed{\rm cc}}}(k_{L},k_{S},k_{S}) =
     -\frac{20}{18\epsilon} \left(\frac{F_Y}{F} \frac{1}{c_L^2 \epsilon}
     + \frac{27}{20} \right)
		P_{\zeta}(k_S)
		P_{\gamma}(k_L)
     \equiv -\frac32 \frac{{\cal R}}{\epsilon}
		P_{\zeta}(k_S)
		P_{\gamma}(k_L),
\end{equation}
the SFSR results in Ref.~\cite{Jeong:2012df} 
are easily adapted to solid inflation.
Ref.~\cite{Jeong:2012df} shows that for SFSR
inflation (and neglecting late-time effects), the smallest
tensor amplitude $A_\gamma$, defined by $P_\gamma(k_L)=A_\gamma
k_L^{-3}$, detectable at the $\gtrsim 3\sigma$ level is $\sim
300 \, (k_{\rm max}/k_{\rm min})^{-3}$, where $k_{\rm min}$ and
$k_{\rm max}$ are the minimum and maximum wavenumbers,
respectively, probed by a given galaxy survey.  Detection of
SFSR tensors near the maximum amplitude, $A_\gamma\simeq 2\times
10^{-9}$, currently allowed requires $k_{\rm max}/k_{\rm min}
\gtrsim5000$, beyond the reach of galaxy surveys but perhaps
within reach of future 21-cm mapping experiments.

In solid inflation there is a similar prediction for this
galaxy four-point correlation function that arises from the part
of the tensor-scalar-scalar bispectrum that satisfies the
consistency condition.  There is then, however, an additional
contribution from the consistency-condition--violating part that
is 
${\cal R}/\epsilon$ times the naive (neglecting late-time effects) prediction given
in Ref.~\cite{Jeong:2012df} from the consistency condition.  This
number must exceed $1/5$ if the cc-violating
four-point signal is to dominate the cc-preserving signal.  If the
tensor-to-scalar ratio is indeed as large as $r\sim 0.1$, then a
model with ${\cal R}/\epsilon \gtrsim 27$ will
give a detectable signal in a galaxy survey, like EUCLID, with
$k_{\rm max}/k_{\rm min} \simeq 750$.  A model with
${\cal R}/\epsilon \gtrsim 1.5$ will
be detectable in a 21-cm survey that maps a volume with
$k_{\rm max}/k_{\rm min} \simeq 5000$.

As seen above, the quadrupole constrains $(H_{c} {\cal R}/\epsilon
M_{P})^2 |\ln (k_L^{min} H_0^{-1})| \lesssim 1$.  Thus, for example, if
$r\sim 0.1$, the quadrupole constraint is (taking the log to be
$\sim10$) ${\cal R}/\epsilon \lesssim 10^{4}$.
We thus see that it is
easily possible to have a solid-inflation model consistent with
the quadrupole constraint and which will still have a
clustering-fossil signature large enough to be detectable in
forthcoming large-scale-structure surveys. In particular,
this range of values for ${\cal R}/\epsilon$ corresponds to a region in the parameter
space of the theory where the parameters naturally lie\footnote{As discussed, the ratio $F_{Y}/F$ needs to be no larger than unity but that leaves plenty of room for an intriguing value of ${\cal R}/\epsilon$ to be accommodated.}. In fact, they may
even be conceivably large enough to be detectable with existing
data!

\section{Conclusions}
\label{sec:conclusion}

In this paper we have studied the squeezed limit of the
tensor-scalar-scalar bispectrum induced during inflation.  We
reviewed how the consistency condition found in
Ref.~\cite{Maldacena:2002vr} relating this bispectrum to the
scalar and tensor power spectra generalizes to any single-clock
inflation model.  We then computed the TSS correlator in 
non-attractor inflation. There the decaying mode (which is ordinarily negligible in single-clock models) causes a departure from the single-clock dynamics. The non-attractor inflationary phase, which is followed by a more traditional attractor phase, leaves an important imprint on the three-point function in the squeezed limit; specifically, it manifests itself at quadratic order in the soft (tensor) momentum. 
In solid inflation, the anisotropic stress of the medium 
leads to a direct violation of the consistency conditions.

The TSS bispectrum in non-attractor inflation can give rise to
an apparent quadrupolar departure from statistical isotropy (SI)
in large-scale structure, with the SI violation most significant
at the largest scales.  The consistency of the CMB quadrupole
with SI constrains the transition from the non-attractor to
attractor phase to occur before the time that the current
observable Universe exited the horizon during inflation.  The
very rapid decay of the SI violation suggests that there will be
no further observable consequences of the squeezed-limit TSS on
smaller scales.

The effects of the TSS bispectrum from solid inflation are
distributed much more evenly among different distance scales.
Thus, it is conceivable that there may be clustering fossils of
the type discussed in Ref.~\cite{Jeong:2012df} in large-scale
structure of a magnitude that could be detectable with
forthcoming survey, and possibly even with current data.
Heuristically, the effects of the anisotropic medium that fills
the Universe during inflation may be written in the distribution
of galaxies today!  We thus encourage the pursuit of such
signatures.

Here we have shown only that these effects may occur with
appreciable magnitudes in solid inflation.  More work must be
done to map out the parameter space of solid-inflation models
(as well as related models, like gauge-flation or chromo-natural
inflation
\cite{Adshead:2012kp,SheikhJabbari:2012qf,Dimastrogiovanni:2012st,Dimastrogiovanni:2012ew,Maleknejad:2012fw})
in which such signatures may arise.  It will also be
interesting to explore the magnitude of effects induced by the
TSS bispectrum in other models of inflation that may violate the
consistency conditions.

\acknowledgments

It is a pleasure to thank Junpu Wang and Lasha Berezhiani for  useful discussions, and Razieh Emami and Hassan Firouzjahi for valuable correspondence and comments on an earlier version of the manuscript.
MK and DJ were supported by NSF Grant No. 0244990 and the John
Templeton Foundation. ED acknowledges partial support from the DOE grant DE-SC0011842 at the University of Minnesota. MF is supported in part by DOE  DE-SC0010600. ED and MF would like to thank the Cosmology group at Johns Hopkins Physics and Astronomy Department for very warm hospitality whilst parts of this work were being completed.

\appendix 
\section{Zeroth and first order CCS}
\label{sec:fieldredef}
For completeness, we report here from \cite{Berezhiani:2013ewa} (see also \cite{Creminelli:2012qr}) the SSS and TSS consistency conditions up to linear order in the soft momentum, which is their most familiar form. For the scalar one has
\bea
\frac{\langle \zeta_{\vec{q}} \zeta_{\vec{p}} \zeta_{-\vec{q} - \vec{p}} \rangle' }{P_\zeta(q)}=\qquad \qquad  \\
-\left(3+p_k\frac{\p}{\p p_k}\right)P_\zeta( p )
&-&\frac{1}{2}q_k\left( 6\frac{\p}{\p p_k}- p_k\frac{\p^2}{\p p_a \p p_a} +2p_a\frac{\p^2}{\p p_a \p p_k} \right)P_\zeta( p )+\mathcal{O}(q^2)\,.  \nonumber
\eea
while the TSS reads
\bea
\frac{\langle \gamma^{ij}_{\vec{q}}\zeta_{\vec{p}} \zeta_{-\vec{q} - \vec{p}} \rangle'}{P_\gamma(q)} =  \quad\qquad \qquad\qquad \qquad\qquad \qquad\qquad \qquad \qquad \qquad\qquad \qquad\qquad \qquad \\ -\frac{1}{2}\hat{P}^{ijk\ell }(\hat{q}) p_k\frac{\p}{\p p_\ell} P_\zeta( p )
 +  \frac{1}{4}\hat{P}^{ijk\ell}(\hat{q}) q_m \left( p_m \frac{\p^2}{\p p_k \p p_\ell}-2p_k \frac{\p^2}{\p p_\ell \p p_m} \right)P_\zeta( p ) +   \mathcal{O}(q^2).  \nonumber
\eea

\section{The tensor-scalar-scalar action}
\label{sec:tssaction}

The action at third order in the perturbations
$\gamma\zeta\zeta$ has the form
\cite{Maldacena:2002vr,Arroja:2008ga,Bartolo:2010bu},
\be\label{in}
     \mathcal{S}_{\gamma\zeta^{2}}=\int dt d^{3}x
     \Big\{-2\frac{a}{H}\gamma_{ij}\partial_{i}
     \dot{\zeta}\partial_{j}\zeta-a\gamma_{ij}
     \partial_{i}\zeta\partial_{j}\zeta-\frac{1}{2}a^{3}\left(3\zeta
     -\frac{\dot{\zeta}}{H}\right)\dot{\gamma}_{ij}
     \partial_{i}\partial_{j}\psi+\frac{1}{2}
     a^{3}\partial_{k}\gamma_{ij}  \partial_{i}\partial_{j}\psi
     \partial_{l}\psi\Big\},
\ee
where $\psi$ represents the shift function of the metric,
Eq.~(\ref{admf}), $N^{i}=\partial_{i}\psi$, with
$\psi=-\zeta/aH+\chi$ and
$\partial^{2}\chi=(\epsilon/c_{s}^{2})\dot{\zeta}$. Multiple
partial integrations can be performed to bring Eq.~(\ref{in})
to a simpler form,
\bea\label{intL}
     \mathcal{S}_{\gamma\zeta^{2}} &= & \int dt d^{3}x \Big\{
     \epsilon a \gamma_{ij}\partial_{i}\zeta\partial_{j}\zeta
     +\frac{1}{4}a^{3}\partial^{2}\gamma_{ij}
     \partial_{i}\chi\partial_{j}\chi +\frac{1}{2}\epsilon
     a^{3}\dot{\gamma}_{ij}\partial_{i}
     \zeta\partial_{j}\chi\nonumber\\ 
     &+&f\left(\zeta,\gamma\right)\frac{\delta\, L}{\delta \,
     \zeta} +f_{ij}\left(\zeta,\gamma\right) \frac{\delta\,
     L}{\delta \, \gamma_{ij}}\Big\}\,],
\eea
where the last two terms are proportional to the equations of
motion for $\zeta$ and $\gamma$ and can be therefore eliminated by a
field redefinition
\begin{equation}
     \zeta=\zeta_{n}+f\left(\zeta_{n},\tilde{\gamma}_{ij}\right),\qquad
     \gamma_{ij}=\tilde{\gamma}_{ij}+f_{ij}
     \left(\zeta_{n},\tilde{\gamma}_{ij}\right).
\end{equation}
The complete expressions for the functions $f$ and $f_{ij}$ are
\cite{Maldacena:2002vr,Jarnhus:2007ia,Arroja:2008ga},
\bea\label{fr1}
     f(\zeta_{n},\tilde{\gamma}_{ij}) &\equiv &\frac{1}{2}
     \frac{\ddot{\phi}}{\dot{\phi}H}\zeta_{n}^{2}
     +\frac{\epsilon}{2} \zeta_{n}^{2}
     +\frac{1}{H}\dot{\zeta}_{n}\zeta_{n} -\frac{1}{4}
     \frac{1}{a^{2}H^{2}}\left(\partial\zeta_{n}\right)^{2}
     +\frac{1}{4}\frac{1}{a^{2}H^{2}}\partial^{-2}
     \partial_{i}\partial_{j}\left(\partial_{i}\zeta_{n}
     \partial_{j}\zeta_{n}\right)\nonumber\\  
     &+&\frac{1}{2}\frac{1}{H}\partial_{i}\chi_{n}
     \partial_{j}\zeta_{n} -\frac{1}{2}\frac{1}{H}
     \partial^{-2}\partial_{i}\partial_{j}
     \left(\partial_{i}\chi_{n}\partial_{j}\zeta_{n}\right)
     -\frac{1}{4}\frac{1}{H}\dot{\tilde{\gamma}}_{ij}
     \partial^{-2}\partial_{i}\partial_{j}\zeta_{n},\\\label{fr2}
     f_{ij}(\zeta_{n},\tilde{\gamma}_{ij}) &\equiv &
     \frac{1}{H}\dot{\tilde{\gamma}}_{ij}\zeta_{n}-\frac{1}{a^{2}H^{2}}
     \partial_{i}\zeta_{n} \partial_{j}\zeta_{n}+\frac{1}{H}
     \left(\partial_{i}\chi_{n}\partial_{j}\zeta_{n}
     +\partial_{j}\chi_{n}\partial_{i}\zeta_{n}\right) .
\eea

We report below the contributions to the tensor-scalar-scalar bispectrum from
the field redefinitions. The
leading-order contributions (in powers of $H/M_{P}$) have the
form,
\be\label{fR}
     \mathcal{B}(k_1,k_2,k_3)_{FR} =
     \mathcal{B}_{[\zeta\rightarrow\tilde{\gamma}\zeta_{n}]}
     (k_{1},k_{2},k_{3})+\mathcal{B}_{[\gamma\rightarrow\zeta_{n}^{2}]}
     (k_{1},k_{2},k_{3}).
\ee
The first contribution in Eq.~(\ref{fR}) arises from
Eq.~(\ref{fr1}), specifically from the expression of the
curvature fluctuations as a scalar-tensor convolution,
\be
     \zeta \rightarrow
     -\frac{1}{4}\frac{1}{H}\dot{\tilde{\gamma}}_{ij}
     \partial^{-2}\partial_{i}\partial_{j}\zeta_{n}. 
\ee
The second contribution is due to Eq.~(\ref{fr2}).  In
particular, it comes from the redefinition of a tensor
fluctuation in terms of the convolution of two scalars,
\bea
     \gamma_{ij}\rightarrow -\frac{1}{a^{2}H^{2}}\partial_{i}
     \zeta_{n}\partial_{j}\zeta_{n}+\frac{1}{H}
     \left(\partial_{i}\chi_{n}
     \partial_{j}\zeta_{n}+\partial_{j}
     \chi_{n}\partial_{i}\zeta_{n}\right).
\eea
In the limit $k_{1}\equiv k_{L}\ll k_{2}\simeq k_{3}\equiv k_{S}$, one finds
\bea
     &&\mathcal{B}_{[\zeta\rightarrow\tilde{\gamma}\zeta_{n}]}
     (k_{L},k_{S},k_{S}) = \frac{1}{8}
     \frac{H_{*}^{4}}{M_{P}^{4}} \left(\frac{1}{\epsilon_{*}
     c_{s}}\right) \left(\frac{1}{k_{S}^{3}k_{L}^{3}}\right)
     \frac{(k_{S}c_{s}\tau_{0})^{2}}{c_{s}^{2}
     }\left(\frac{k_{L}}{k_{S}}\right)^{2},\\
     &&\mathcal{B}_{[\gamma\rightarrow\zeta_{n}^{2}]}
     (k_{L},k_{S},k_{S})=
     \frac{1}{8}
     \frac{H_{*}^{4}}{M_{P}^{4}} \left(\frac{1}{\epsilon_{*}
     c_{s}^{}}\right) \left(\frac{1}{k_{S}^{3}k_{L}^{3}}\right)
     \left(\frac{1}{\epsilon_{*}}-2\right)
     \frac{(k_{S}c_{s}\tau_{0})^{2}}{c_{s}^{3}}
     \left(\frac{k_{L}}{k_{S}}\right)^{3}, \nonumber\\
\eea
where $\tau_{0}$ is the time of observation.

\end{document}